\documentclass[structabstract]{aa}  
\usepackage{graphicx}
\usepackage{txfonts}
\usepackage{array}
\usepackage{longtable}
\usepackage{multirow}

\usepackage{natbib}

\begin{document}
   \title{Bright Blazhko RRab Lyrae stars observed by ASAS and the SuperWASP surveys}


   \author{M. Skarka
          }

   \institute{Department of Theoretical Physics and Astrophysics, Faculty of science, Masaryk University, 
   						Kotl\'a\v rsk\'a 2, Brno, Czech Republic\\
              \email{maska@physics.muni.cz}
             }


 \abstract
{}
{Period study of 321 fundamental mode RR Lyrae type stars (RRab), which had appropriate data in ASAS and SuperWASP surveys, was performed to complement and extend the list of known Blazhko stars in galactic field with bright stars up to 12.5 mag in maximum light.}
{An individual approach was applied to each studied star. Permanent visual supervision was maintained to each procedure in data analysis (data cleaning, frequency spectra examination) to avoid missing any possible sign of the Blazhko effect. Period analysis was performed using \textsc{Period04} software.}
{We found 100 stars to be definitely modulated. In 25 cases, previously unknown modulation was revealed and 8 new candidates for Blazhko stars were identified. Their modulation needs to be confirmed. In 18 previously found Blazhko stars, no modulation was detectable. Multiple modulation was revealed for eight stars that were previously proposed to show simple modulation. In total, there were twelve stars with some peculiarity in their modulation in the sample. This brings the incidence rate of multiple/irregularly-modulated stars to 12~\%. The ratio of the modulation periods of five of the double-modulated stars was within the ratios of small integers. One of stars studied, IK Hya, showed a very interesting frequency spectrum, which we interpret as changing Blazhko period between 71.81 and 75.57~days and an additional 1403-day-long cycle analogous to a four-year cycle of the prototype RR Lyr. The limits of the shorter period produce a beating period that is approximately twice as big as a 1403-day period. The newly revealed Blazhko star RZ~CVn seems to undergo changes in the amplitude of the modulation, as well as in the basic pulsation and Blazhko periods. We found that the incidence rate of the Blazhko RR Lyraes is at least 31~\%, more likely even higher. It was also found that the majority of the Blazhko variables show triplet structures in their frequency spectra and that in 89~\% of these cases, the peak with larger amplitude is on the right-hand side of the main pulsation component.}
{}

	
	 \authorrunning{M. Skarka}
	 \titlerunning{Bright Blazhko stars from ASAS and SuperWASP surveys}
	
   \maketitle
%

\section{Introduction}

Data from automatic sky surveys provide an invaluable opportunity for finding new variables and for long-term monitoring of the sky. The most extensive survey of Galactic field variable star research are without doubt Polish ASAS \citep[\textit{All Sky Automated Survey}; see e.g.][]{pojmanski1997,pojmanski2002} and SuperWASP \citep[\textit{Super Wide Angle Search for Planets}, hereafter SW; ][]{pollaco2006,butters2010} managed by the University of Leicester. Both projects have been run for several years.

Field RR Lyraes have been studied many times on the basis of various sky surveys. Among others, \citet{kinemuchi2006} and \citet{wils2006} utilized data from \textit{Northern sky variability survey} \citep[NSVS, ][]{wozniak2004}, e.g. \citet{kovacs2005}, \citet{szczygiel2009} 
studied RR~Lyraes based on ASAS data, and \citet{drake2013} dealt with RR Lyraes measured in \textit{Catalina sky survey}\footnote
{http://www.lpl.arizona.edu/css}. Probably the most precise ground-based survey data were provided by \textit {the Konkoly Blazhko survey}, which have been operated since 2004 \citep[e.g.][]{jurcsik2009a,sodor2007,sodor2012}.

Crucial advances in Blazhko effect research have been made recently by space missions \textit{CoRoT} and \textit{Kepler}. Ultra-precise data provided by these space telescopes allowed findings of, say, cycle-to-cycle variations now known as period doubling \citep{szabo2010,kolenberg2010,benko2010} or long-term changes in the Blazhko effect and excitation of additional modes \citep{poretti2010,chadid2010,guggenberger2012}.

Large samples of Blazhko RR Lyrae stars in galactic bulge were also studied in the framework of MACHO and OGLE surveys 
\citep{soszynski2011, moskalik2003}. Since the stars in the bulge are faint and are not included in ASAS or SW surveys, we did not deal with them.

Special papers based on ASAS data, which are at least partially devoted to Blazhko stars in our Galaxy, were published by \citet{kovacs2005}, \citet{wils2005} (hereafter WS05), and \citet{szczygiel2007} (hereafter SF07), who all revealed many new Blazhko variables and gave a list of them. An extensive study of the Blazhko effect is presented in the paper of \citet{leborgne2012}, which is based on observations made with \textit{TAROT} telescopes \citep{klotz2008,klotz2009}. \citet{leborgne2012} also made use of $O-C$ information stored in the \textit{GEOS RR Lyr database} \citep{leborgne2007} and gave a list of galactic field Blazhko stars. A regularly updated database of known Blazhko RR Lyraes situated in the galactic field \citep[\textit{BlaSGalF}, ][]{skarka2013b}\footnote{http://physics.muni.cz/$\sim$blasgalf/} recently contained about 270 objects.

There are a few studies of particular RR Lyraes proceeded from SW mesurements, e.g. the TV Boo study made by \citet{skarka2013a}, or 
\citet{srdoc2012}, who revealed the Blazhko effect in GSC02626-00896, but a comprehensive study of RR Lyraes based on SW data has not been 
performed yet. 

RR Lyraes with appropriate ASAS and SW data were in the spotlight. The goal was to search for the presence of the Blazhko modulation and to describe its characteristics (pulsation and modulation periods and multiplicity/irregularity of the modulation). We applied commonly used techniques of light curve and frequency spectra examination to complement and extend the list of known Blazhko stars with bright RR Lyraes (up to 12.5 mag in maximum light). To be as precise as possible, we carefully cleaned and analysed each light curve and frequency spectra individually.

\section{The data}

\subsection{Data characteristics}

Both ASAS and SW data are products of measurements with low-aperture telescopes. ASAS uses Johnsons $V$ and $I$ filters, while SW telescopes are equipped with a broad-band filter (from 400 to 700 nm) \citep{pojmanski2001,pollaco2006}. The typical scatter of points is difficult to define, because it changes from target to target. In the case of ASAS it is from about 0.02 to 0.1~mag, while SW~scatter extends from 0.007 to 0.04~mag for our sample stars. Also the time span and the distribution of the data differ for both surveys and various targets.
	
\begin{figure}[htbp]
   \centering
   \includegraphics[width=9cm]{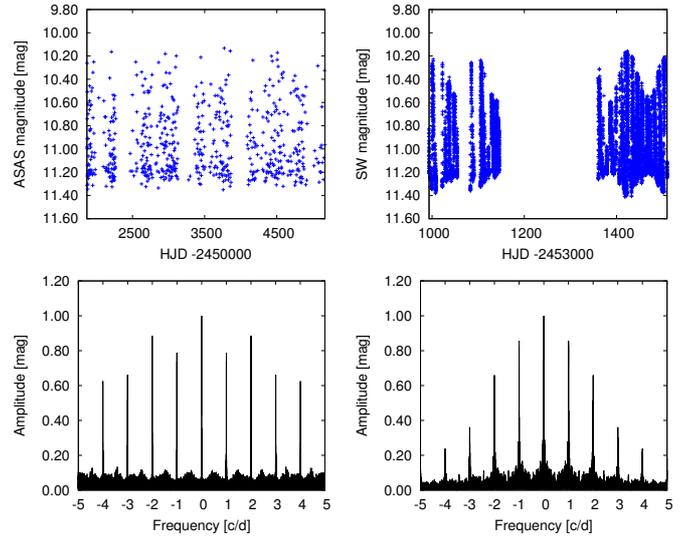}
   \caption{Differences between ASAS (left column) and SW (right column) data of RY Col together with their spectral windows (bottom panels). ASAS data are more or less uniformly distributed, while SW data show a large gap between observing seasons. Both spectral windows are dominated by strong daily aliases. Except these, SW data suffer from additional artefacts, namely from yearly aliases.}%
   \label{specwin}
   \end{figure}

ASAS data typically cover a few years (up to year 2009) with point-to-point spacing between two and four days with only small gaps between observational seasons. Typical sampling of SW data points is between five and twelve minutes. Therefore SW data are generally more numerous and more dense than ASAS data, but they typically span only one or two seasons\footnote{with few exceptions, which cover three years} (see last columns in Tables 1-5). These characteristics resulted in a typical Nyquist frequency from 0.17 to a few tens for ASAS and in 100 to more than 1000~c/d in the case of SW. For differences between ASAS and SW data see fig. \ref{specwin}, where data for RY~Col are plotted with their spectral windows. 

\subsection{Sample selection and data cleaning}

The basic set of the stars originated in the GCVS catalog \citep{samus2012}, which contains 512 stars of RRab type brighter than 12.5~mag in maximum. This magnitude limit was chosen to analyse only stars with the best data quality -- the scattering of data points over 13~mag rapidly increases. The list extracted from GCVS was completed with 82 stars, which did not have a GCVS designation in 2012 and which were taken from SF07 and \citet{szczygiel2009}. There were data for 475 stars available in the ASAS database, and for 243 RR Lyrae stars in the SW archive. Data of 161 stars appeared in both the ASAS and SW archives. This means that a total of 557 stars remained for further analysis.

Unfortunately, data from automatic surveys are often very noisy and of bad quality. Therefore we first looked for ephemeris 
of the stars in the \textit{Variable star index}, VSX \citep{vsx} to be able to construct phased light curves. This was done to quickly check the quality of the data. In many cases, the periods that we found were not current. Therefore we were forced to update and roughly estimate the actual pulsation period. For this purpose, as for all our other period analyses, \textsc{Period04} software \citep{lenz2004} was used. 	
		
In the case of ASAS survey, only the best quality $V$-data (with flags A and B) were used. Following \citet{kovacs2005}, ASAS $V$ magnitude was calculated as the weighted average of five values in different apertures using	
	\begin{equation}
		V(t)=\frac{\sum^{5}_{i=1} V_{i}\left(t\right)\sigma^{-2}_{i}\left(t\right)}{\sum^{5}_{i=1}\sigma^{-2}_{i}\left(t\right)}.	
	\end{equation}	
Only light curves with more than 200 points from ASAS and with more than 1000 points from SW were left in the sample for further investigation.  	
     
\begin{figure}
   \centering
   \includegraphics[width=7.6cm]{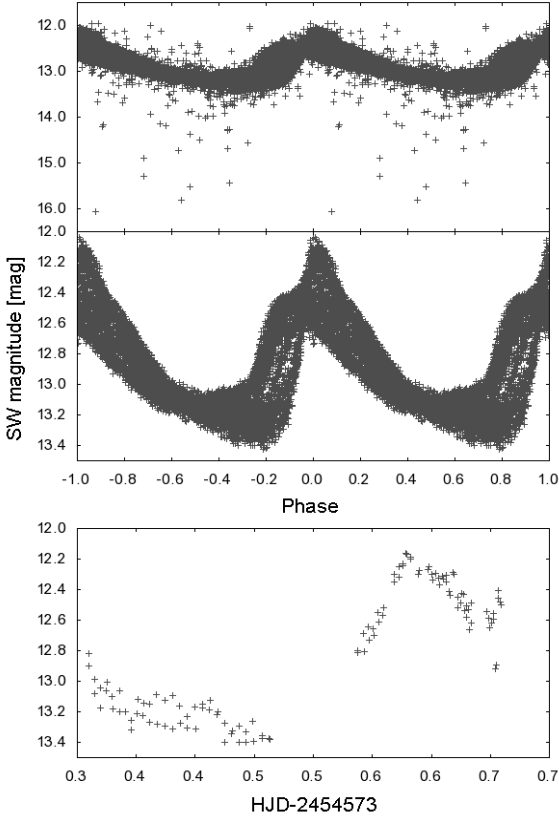}
   \caption{Steps in the cleaning process in V559 Hya. There are raw SW data folded with the main pulsation period taken from 							VSX on the top panel. The middle panel shows phased data after cleaning process. The bottom panel is a sample of one of 							omitted nights, which was discarded due to poor quality of the data.}
   \label{cleaning}
\end{figure}

Once the phased light curves were prepared, we briefly surveyed them to immediately discard stars whose data, at a glance, were very noisy. Remaining light curves were then fitted with the sum of sines and points that were farther away than 0.5~mag from the fit were automatically removed. This procedure guaranteed that real outliers were eliminated and points scattered by a possible Blazhko effect remained in the dataset. A careful visual examination of all light curves was then applied, and points, which were unambiguously caused by noise, were manually deleted. In the case of SW stars, we also visually scanned time series night-by-night and removed outliers, as well as entire noisy nights. Figure \ref{cleaning} shows steps in the cleaning process in the case of V559 Hya.

\begin{figure}
   \centering
   \includegraphics[width=8.5cm]{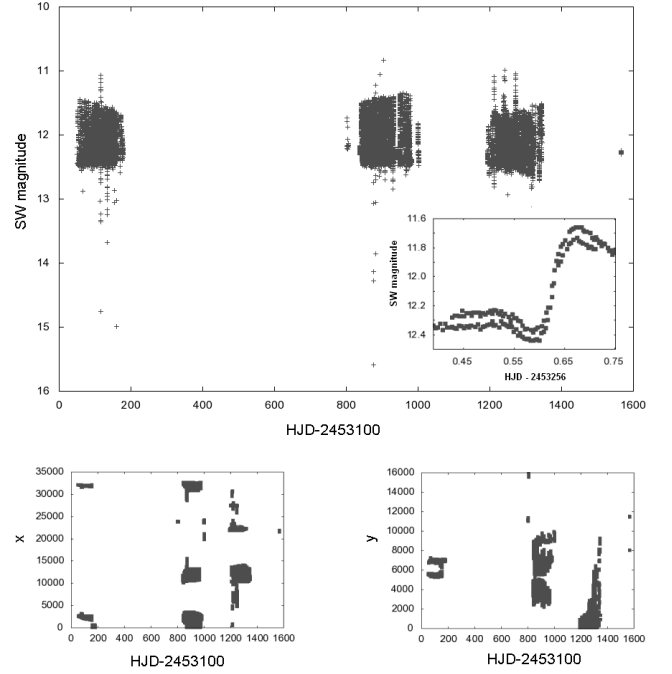}
      \caption{Illustration of difficulties in SW data processing on the case of DM~And. The upper panel shows raw data. 
      				 The change in amplitude of this star caused by the Blazko effect is apparent. The detail shows 
      				 the night HJD 2453256, where the vertical shift between data coming from various cameras is demonstrated. 
      				 The distribution of the positions of the star on the chip in x and y directions during the whole observation 
      				 period are shown in the bottom part of this figure.}
         \label{SWdifficulties}
\end{figure}

SW data often contained vertically shifted multiple points. This was the consequence of simultaneous imaging of the target by a few cameras. To avoid using such data, the information about the position of the star on the chip (fig. \ref{SWdifficulties} bottom panels) was utilized. If the separation of the data from two cameras was impossible, each star with a vertical separation of points larger than 0.01 mag was discarded. This was the case, for instance, for DM And (fig. \ref{SWdifficulties}), which shows previously unknown modulation. 

Several targets from SW underwent vertical shifts of points taken in different seasons (as the third season of DM Cyg, fig. \ref{SWdifficulties}). 
If so, seasons were phased separately, and the light curves were fitted with the sum of sines to get zero points. Afterwards, the 
points from different seasons were shifted to fit one each other. This procedure was used in the case of U Tri, among others, and magnitudes of the first part of the dataset were decreased by 0.09~mag. In some cases, vertical shifts were present even in one season. In such cases, we removed the whole season from the dataset. If all seasons underwent shifts, the star was discarded from our sample. Only 321 variables met all the above-mentioned criteria and were accepted for further analysis.


\section{Light curve and frequency spectra examination -- looking for the Blazhko effect}

	\subsection{Introductory remarks on the Blazhko effect}

The manifestations of the Blazhko effect\footnote{Named after one of its discoverers S.N.\citet{blazhko1907}, who noticed periodic changes in the maximum timings of RW Dra.}, which is usually defined as phase and/or amplitude modulation of the light curve with periods between a few days and several thousand days, can be very heterogeneous and in many cases are difficult to uncover. The most conspicuous feature of the Blazhko effect is the change in amplitude of the light variations, which can be more than 0.5~mag \citep[e.g. XY Eri or IK Hya, ][]{leborgne2012}, but also less than a tenth of a magnitude \citep[e.g. DM Cyg, ][]{jurcsik2009b} or even only a few hundredths of magnitude -- KIC 11125706 with the amplitude of the modulation of only 0.015~mag \citep{benko2010}. The same applies to the range of phase variations -- in RR~Gem the Blazhko effect causes scarcely perceptible changes in $O-C$ of 0.004~d \citep{leborgne2012}, while some stars change their $O-C$s in more than 0.07~d \citep[e.g. XY Eri, V445 Lyr ($\Delta O-C=0.15$~d) ][]{leborgne2012,guggenberger2012}.			
		 																	
When analysing the frequency spectra of Blazhko stars, we can expect characteristic structures to be present. Firstly, the Blazhko effect manifests itself as an additional peak $f_{\mathrm{BL}}$ in the range of $\sim$0.0001 to 0.2~cycles per day\footnote{This peak is present only if the star undergoes amplitude modulation, which satisfy the vast majority of the Blazhko RR~Lyraes \citep{benko2011}.}. But, the most prominent features of Blazhko-frequency spectra are characteristic series of equidistant peaks around basic pulsation frequency and its harmonics ($kf_{0}\pm lf_{\mathrm{BL}}$, where $f_{0}$ is the basic pulsation frequency and $k$ and $l$  are integers) with rapidly decreasing amplitudes. In data with limited precision, multiplets can be strongly reduced by noise and only the highest peaks are found. In such a case, we observe only triplets or even doublets. In contrast to limited ground-base observations, in data gathered from space, high-order multiplets are observed \citep[e.g. in V1127 Aql, where sepdecaplets ($l=8$) were identified][]{chadid2010}.
		
Multiplets must not be strictly equidistant and side peaks do not have to have the same amplitude -- it depends on many factors, 
mainly on the mixture of amplitude and frequency modulation \citep[for more details see][]{benko2011}. In addition, recent studies have shown that the Blazhko effect can be multiple \citep[e.g. like CZ Lac ][]{sodor2011}, a very irregular and complex phenomenon affected by long-term and/or sudden changes, as in RR Lyrae itself, where a four-year long cycle was found, during which pulsation and modulation characteristics are changing \citep{detre1973,kolenberg2011b,stellingwerf2013}. During this long cycle, the Blazhko effect can almost disappear. 	
		
Another examples of complexity of modulation show CoRoT 105288363 and V445~Lyr, which seem to undergo drastic changes in their 
Blazhko effect \citep{guggenberger2011,guggenberger2012}. Changing Blazhko modulations like these, especially strong ones like in V445~Lyr, will lead to a different frequency pattern with more peaks than the normal multiplet. There will be additional peaks or even additional multiplets in cases with an irregular or changing modulation.
		
All these illustrations demonstrate the variety and complexity of the Blazhko effect. Especially when analysing data from ground-base automated surveys, one should be very careful when identifying the Blazhko effect and interpreting the frequency spectra. It is very difficult to distinguish between long-term Blazhko modulation, period changes and multiple or irregular modulation, because all these processes produce additional peaks close to the basic pulsation components. If the data are of bad quality or are badly distributed or if the observations are not sufficiently spread out in time, the presence of the Blazhko effect in frequency spectra could even stay hidden.
		
Although astronomers have known about the existence of the Blazhko effect since the beginning of the 20th century and much progress have been made \citep[e.g. revealing period doubling by ][]{szabo2010}, a suitable physical explanation is still not available. For more information about the Blazhko effect and proposed models in general see e.g. \citet{kovacs2009}, \citet{kolenberg2011a}, and references therein. 

   \begin{figure*}[htbp]
   \centering
   \includegraphics[width=18cm]{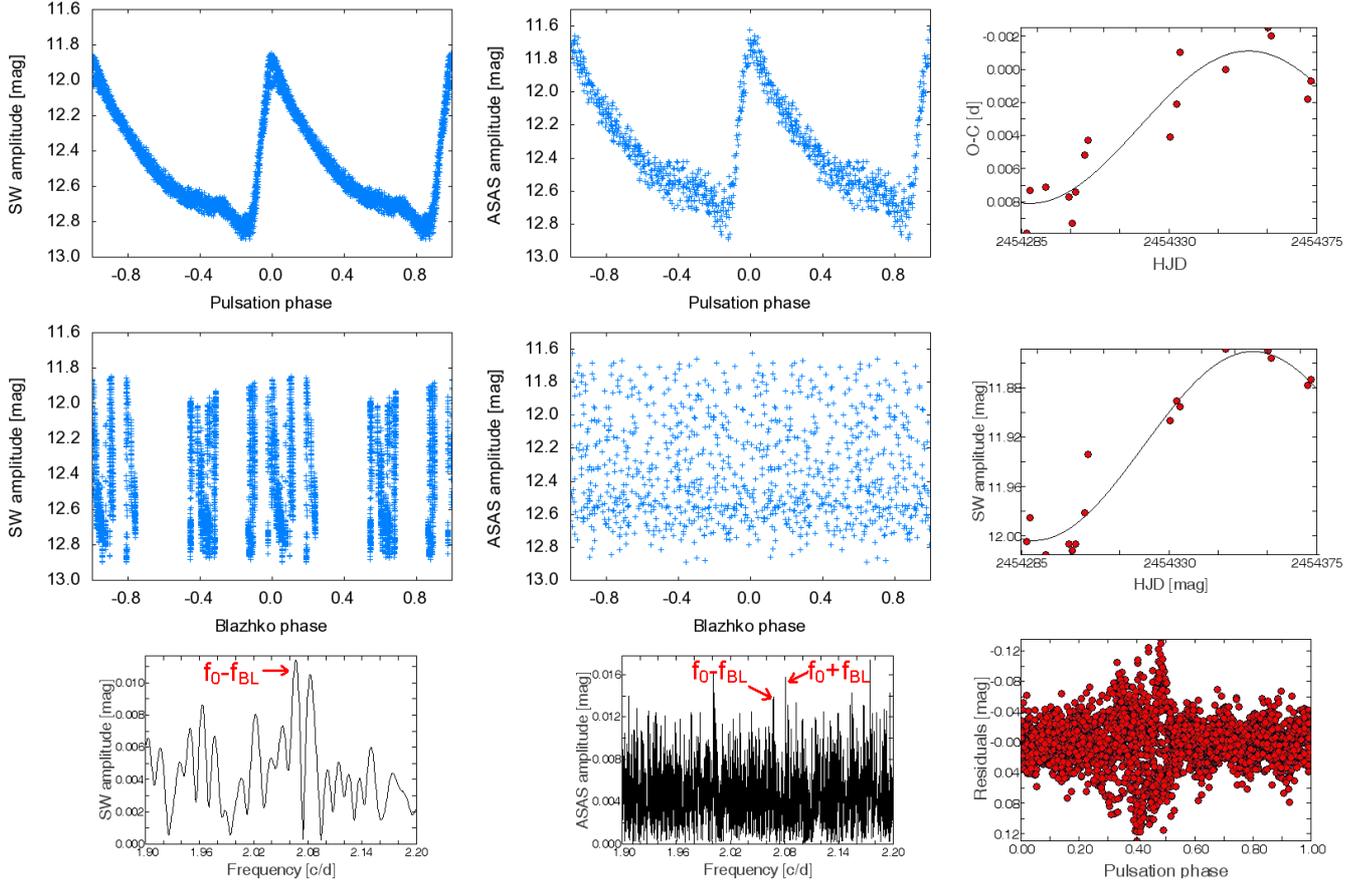}
   \caption{Analysis of HH Tel. First column shows cleaned SW data folded according to epoch 2454347.2643 with the main 
   					pulsation period $P=0.4820925$~d (top panel) and with the Blazhko period 133~d (middle panel). Frequency spectrum 
   					based on SW data in the vicinity of $f_{0}$ prewhitened with $kf_{0}$ is shown in the bottom panel in the first 
   					column. Second column displays the same as the first one but for the ASAS data. In the third column (based on SW 
   					data again), there are from the top changes in $O-C$ values, changes of maximum brightness, and residuals after prewhitening with the main pulsation components at the bottom (epoch is shifted of a half of a period 
   					for better arrangement). For details see the text in section 3.2.}%
   			\label{methods}
   \end{figure*}	

	\subsection{Methods of searching for the Blazhko effect}

Searching for the Blazhko effect actually started during the cleaning process when phased light curves were examined. High-amplitude modulation was easily resolvable by a fleeting glance at the data. Our methods, inspired by methods used e.g. in \citet{alcock2003}, WS05, and \citet{leborgne2012}, can be summarized in four points:		
		\begin{enumerate}
		\item {\it Phased light curve visual examination} --  visual inspection of the phased light curve to reveal a characteristic 
		scattering around maximum light. 
		\item {\it Frequency spectra analysis with }\textsc{Period04} -- frequencies were fitted 
		independently; i.e., we scanned for frequencies with the highest amplitude and subsequently subtracted them. This was done 
		in the range of $0-30$~c/d down to the signal-to-noise ratio $(S/N)$ of the peaks higher than four\footnote{This limit is usually 
		used for plausible analysis \citep{breger1993}}. We searched for the manifestation of the Blazhko effect in the vicinity of 
		the main pulsation components, as well as for the presence of the Blazhko peak itself (in the range of $\sim$0.0001 to 
		0.2~c/d). All uncertainties were estimated using \textsc{Period04} tools. 
		\item {\it Checking of residuals} -- In stars with only basic pulsation components in frequency spectrum, residuals after 
		prewhitening were visually inspected to see whether there were any typical inhomogeneities possibly caused by the Blazhko 
		effect.
		\item {\it Magnitude-at-the-maximum-light vs. time-of-the-maximum series analysis} -- this was done only for the SW sample, 
		because ASAS light curves, with few exceptions, did not contain continuous time series to determine times and magnitudes of 
		the maximums. For $Mag_{\mathrm{max}}$ and $T_{\mathrm{max}}$ determination, low-degree polynomial fitting (typically up to 
		5th order) was used. Only well defined maximums were fitted. In the next step, frequency analysis of such a series was 
		performed to search for significant peak in the range of $\sim$0.0001 to 0.2~c/d. 
		\end{enumerate} 

Each star was analysed on an individual basis without any automatic procedures (except outliers removing and prewhitening of the frequency spectra). We maintained permanent visual supervision of all steps to avoid omitting any possible sign of the Blazhko effect. A candidate was marked as a Blazhko star only if we were able to find the modulation period or if there was no doubt about the Blazhko effect. Some exceptions, e.g. LS Boo, are discussed in sects. 4.3 and 5.1. 
		
As an example of the procedures used, the analysis of HH~Tel (fig. \ref{methods}) is demonstrated. This star was revealed to be a new Blazhko star. Data from both surveys are plotted at the top in the first two columns of fig. 4. We can see a weak sign of the modulation around maximum light (especially in SW data). This indicated that the Blazhko effect might be present. Frequency spectra analysis down to $S/N=4$ unveiled only the main pulsation components. After prewhitening peaks that could possibly be caused by the Blazhko effect ($f_{0}\pm f_{\mathrm{Bl}}$) were identified, but they are of very low amplitude with $S/N\sim3$ (fig. 4 two left bottom panels). No distinct peak in the range of $\sim$0.0001 to 0.2~c/d was noted. 
		
Therefore, searching for additional peaks in the frequency spectra of the light curve had ambiguous success. But the SW residuals (right bottom corner of fig. 4) showed a tell-tale characteristic of the Blazhko effect: typically scattered points around the position corresponding to ascending branch and maximum light.

The last step was to determine the modulation period. For this purpose, the times ($T_{\mathrm{max}}$) and brightnesses of 
maximum light ($Mag_{\mathrm{max}}$) were used. Period analysis of such data directly revealed a modulation period of 133(9)~d, while weak peaks identified in Fourier spectra of the light curves gave values 143(1)~d and 123(5)~d for ASAS and SW data, respectively. All values are very inaccurate, which is due to the limited time scale of SW data (only one Blazhko cycle) and due to the imprecision of ASAS data.
		
At that stage we were able to construct Blazhko phased light curves (first two columns in the middle of fig. 4) to check for the correctness of the value we found. SW data showed a typical \textquoteleft Blazhko\textquoteright light curve, while there was no sign of modulation in the ASAS Blazhko light curve\footnote{Compare with the light curve in fig. 9, where the ASAS data are phased with 143-day period and where the Blazhko effect is apparent.}. 
		
HH Tel was chosen intentionally to show that it is very problematic to decide the Blazhko nature of this star based on ASAS data 
alone, while there is no doubt about the modulation in SW data. The example of HH Tel clearly shows that many unveiled Blazhko stars could remain in the ASAS database.


\section{Blazhko stars from ASAS}

	\subsection{Known Blazhko stars}
	
A review of 62 previously known Blazhko stars contained in our ASAS sample is given in Table \ref{table:1}. For comparison the values of Blazhko periods found by WS05 and SF07 are also given (5th and 6th column). These are naturally very similar to ours, because we studied the same, but more extended data. Therefore, with respect to periods, only a revision of the ASAS data was actually performed.
	
Epochs given in Table \ref{table:1} are not times of maxima, which are usually given, but they are arbitrarily chosen to obtain Blazhko-phased light curves, as well as the light curves phased with the main pulsation period, with the maximum around the zero phase. The amplitudes $A_{+}$, $A_{-}$ of the side peaks $f_{\pm} =f_{0}\pm f_{\mathrm{BL}}$ corresponding to the Blazhko period $P_{\mathrm{BL}}$ are in 7th and 8th columns and their ratio ($A_{+}/A_{-}$) in 9th column, respectively. All given Blazhko periods in the study correspond to the side peak with the highest amplitude in frequency spectrum. Label \textquoteleft $n$\textquoteright denotes the number of points used and \textquoteleft $TS$\textquoteright their time span.
	
\begin{figure}[htbp]
   \centering
   \includegraphics[width=9cm]{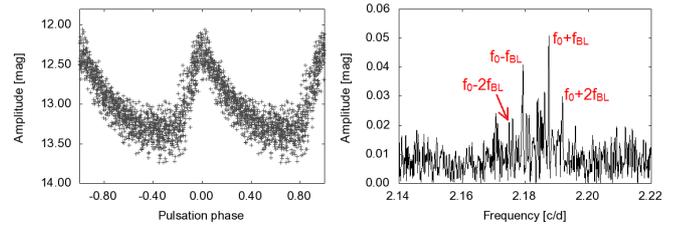}
   \caption{RS Oct -- the only star in our ASAS sample, which showed equally spaced quintuplet.}%
   \label{RSOct}
   \end{figure}

Eight of these stars (marked by an asterisk) showed additional peaks close to the basic pulsation components, which could possibly point to irregularity in modulation, period changes, or multiplicity of a modulation. The modulation periods for VX Aps, GS Hya, and AS Vir were estimated for the first time. The only equally spaced quintuplet was identified in RS Oct (fig. \ref{RSOct}).
	
\def\arraystretch{1}
\newcolumntype{M}{>\tiny l} 
\setlength\tabcolsep{1ex}
	
\begin{table*}
 	\caption{Known Blazhko stars from our ASAS sample}               
	\label{table:1}      
	\centering  
    
	\begin{tabular}{M M M M M M M M M M M}   
	\hline       

Star &	Epoch HJD &	$P_{\mathrm{puls}}$ [d]	& $P_{\mathrm{BL}}$ [d]		& WS05 &	 SF07 &	$A_{-}$ [mag]&	$A_{+}$ [mag]	&$A_{+}/A_{-}$&	 $n$	&	$TS$ [d] \\
\hline
BS	Aps		&	2454527.8757	&	0.5825596(7)	&	46.53(7)	&		&		&	-	&	0.026	&	-	&	646	&	3210	\\
VX	Aps		&	2451981.7844	&	0.484663(2)	&	171.8(6)	&		&		&	0.024	&	0.039	&	1.60	&	647	&	3210	\\
TY	Aps		&	2452088.5376	&	0.501698(1)	&	108.6(2)	&		&		&	0.033	&	0.049	&	1.48	&	601	&	3208	\\
V360	Aqr	&	2452168.6700	&	0.626946(3)	&	54.44(4)	&		&	54.52	&	-	&	0.097	&	-	&	413	&	3254	\\
S	Ara *		&	2454247.7025	&	0.451849(1)	&	49.51(5)	&	49.5	&	49.37	&	0.043	&	0.044	&	1.02	&	507	&	3175	\\
TT	Cnc		&	2454202.5461	&	0.563454(1)	&	88.8(2)	&		&		&	0.022	&	0.045	&	2.05	&	342	&	2546	\\
RV	Cap		&	2452002.1370	&	0.4477426(6)	&	232.3(5)	&		&	231.66	&	-	&	0.105	&	-	&	543	&	3117	\\
Star 1		&	2452034.5590	&	0.526235(1)	&	694(10)	&		&	666.44	&	0.037	&	0.065	&	1.78	&	509	&	3245	\\
BI	Cen		&	2454932.6911	&	0.4531949(4)	&	79.45(9)	&		&	79.01	&	0.039	&	0.042	&	1.08	&	767	&	3179	\\
RX	Cet		&	2454666.8472	&	0.573741(1)	&	261.5(8)	&	256	&	255.5	&	0.058	&	-	&	-	&	397	&	3287	\\
RV	Cet		&	2452080.9219	&	0.6234136(8)	&	112.0(1)	&	112	&	112.05	&	0.025	&	0.040	&	1.60	&	394	&	3275	\\
RX	Col		&	2454751.8312	&	0.593749(3)	&	132.5(4)	&	130	&		&	-	&	0.043	&	-	&	675	&	3298	\\
RY	Col		&	2453854.5121	&	0.4788357(5)	&	82.12(7)	&	82	&	81.95	&	0.028	&	0.073	&	2.61	&	624	&	3298	\\
WW	CrA		&	2452514.1932	&	0.559490(1)	&	35.69(2)	&	35.5	&		&	0.039	&	0.025	&	0.64	&	530	&	3176	\\
V413	CrA	&	2454508.8898	&	0.5893427(4)	&	60.5(2)	&		&	59.96	&	0.009	&	0.010	&	1.11	&	542	&	3158	\\
X	Crt			&	2453449.7158	&	0.732836(1)	&	143.8(3)	&	143	&		&	-	&	0.017	&	-	&	436	&	3296	\\
VW	Dor *	&	2454136.7779	&	0.5705770(7)	&	25.92(2)	&	25.9	&		&	0.014	&	-	&	-	&	667	&	3295	\\
XY	Eri		&	2452553.7594	&	0.554250(1)	&	50.11(5)	&	50	&		&	0.052	&	0.079	&	1.52	&	511	&	3287	\\
LR	Eri *	&	2453408.6223	&	0.602228(1)	&	123.5(6)	&	122	&		&	-	&	0.045	&	-	&	893	&	3296	\\
Star 2		&	2453745.5476	&	0.62963274(9)	&	330(2)	&		&	335.29	&	0.033	&	0.050	&	1.51	&	683	&	3296	\\
RX	For		&	2452082.9124	&	0.5973129(8)	&	31.80(6)	&	31.8	&	31.81	&	-	&	0.061	&	-	&	517	&	3287	\\
SS	For		&	2451869.5957	&	0.4954329(4)	&	34.84(1)	&	34.8	&	34.88	&	0.026	&	0.064	&	2.51	&	501	&	3295	\\
RT	Gru		&	2452858.6870	&	0.512169(1)	&	86.94(8)	&		&		&	-	&	0.079	&	-	&	496	&	3256	\\
DL	Her		&	2453477.7851	&	0.591631(1)	&	34.66(4)	&		&		&	-	&	0.038	&	-	&	427	&	2396	\\
BD	Her		&	2453496.8160	&	0.473794(1)	&	21.68(1)	&		&		&	0.052	&	0.085	&	1.63	&	308	&	2417	\\
GS	Hya		&	2451998.7258	&	0.522829(1)	&	42.8(5)	&		&		&	0.051	&	-	&	-	&	453	&	3186	\\
UU	Hya		&	2453761.7195	&	0.5238652(9)	&	39.93(4)	&		&		&	0.042	&	-	&	-	&	349	&	3112	\\
SV	Hya		&	2452062.5225	&	0.4785476(3)	&	63.23(4)	&	63	&	63.29	&	0.022	&	0.028	&	1.27	&	877	&	3178	\\
V552	Hya	&	2454931.6746	&	0.3987697(8)	&	48.53(3)	&	48.3	&		&	0.142	&	0.072	&	0.51	&	443	&	3177	\\
V559	Hya *	&	2454199.5359	&	0.447944(2)	&	26.469(6)	&	26.6	&	26.5	&	0.076	&	0.135	&	1.78	&	454	&	3187	\\
SZ	Hya		&	2454258.5213	&	0.537225(1)	&	26.26(2)	&	26.3	&	26.3	&	0.076	&	0.080	&	1.05	&	543	&	3298	\\
DD	Hya		&	2454408.8593	&	0.5017706(6)	&	34.44(4)	&		&		&	0.036	&	0.027	&	0.75	&	634	&	3142	\\
SZ	Leo		&	2455012.4827	&	0.534080(9)	&	177.8(7)	&	179	&		&	0.091	&	0.062	&	0.61	&	339	&	2560	\\
BB	Lep		&	2452128.9235	&	0.5389127(9)	&	22.86(1)	&		&	22.84	&	0.039	&	0.041	&	1.05	&	567	&	3300	\\
MR	Lib		&	2453133.7052	&	0.540065(1)	&	41.85(3)	&	41.7	&	41.77	&	0.051	&	0.115	&	2.25	&	430	&	3172	\\
PQ	Lup		&	2453861.6802	&	0.581994(1)	&	48.77(3)	&	48.8	&	48.82	&	-	&	0.068	&	-	&	478	&	3189	\\
V339	Lup	&	2453897.6107	&	0.6005772(7)	&	49.7(9)	&	49.5	&		&	0.014	&	0.029	&	2.07	&	526	&	3191	\\
Star 3		&	2453428.6223	&	0.5501319(9)	&	37.2(2)	&		&	37	&	0.046	&	0.059	&	1.28	&	466	&	3186	\\
UV	Oct		&	2451976.8227	&	0.5425783(5)	&	144.6(2)	&	145	&	143.73	&	0.047	&	0.052	&	1.10	&	1292	&	3233	\\
SS	Oct		&	2453882.8524	&	0.6218497(4)	&	145.0(5)	&	145	&	144.12	&	0.020	&	0.021	&	1.05	&	1401	&	3297	\\
RS	Oct		&	2454255.6649	&	0.458007(1)	&	241(1)	&	244	&	244.2	&	0.045	&	0.058	&	1.29	&	1258	&	3295	\\
DZ	Oct		&	2453055.6301	&	0.4778589(6)	&	36.63(2)	&	36.8	&	36.79	&	0.066	&	0.047	&	1.40	&	1119	&	3294	\\
V2709	Oph	&	2452454.6100	&	0.4613691(7)	&	22.18(1)	&	22.2	&	22.18	&	-	&	0.022	&	-	&	447	&	2944	\\
FO	Pav		&	2452947.5986	&	0.5514365(9)	&	585(6)	&	571	&	557.17	&	0.048	&	0.052	&	1.08	&	536	&	3272	\\
BH	Pav *	&	2452141.8540	&	0.4769604(5)	&	173.7(3)	&		&	173.7	&	0.061	&	-	&	-	&	1038	&	3161	\\
BH	Peg *	&	2452813.8617	&	0.640987(1)	&	175.7(7)	&		&		&	-	&	0.022	&	-	&	213	&	2346	\\
ST	Pic		&	2455105.8195	&	0.485744(1)	&	117.7(1)	&		&	117.9	&	0.011	&	-	&	-	&	660	&	3294	\\
RY	Psc		&	2452944.7300	&	0.529739(1)	&	153.3(5)	&		&	154.53	&	-	&	0.053	&	-	&	372	&	3279	\\
X	Ret *		&	2453823.5050	&	0.491995(2)	&	160.3(3)	&	161	&	160.64	&	0.039	&	0.052	&	1.33	&	1081	&	3273	\\
V1645	Sgr*	&	2452141.9570	&	0.5529477(9)	&	1302(12)	&		&	1331.74	&	0.053	&	0.081	&	1.53	&	747	&	3271	\\
V2239	Sgr	&	2452031.8445	&	0.4401917(8)	&	45.24(6)	&		&	45.39	&	0.040	&	0.051	&	1.26	&	386	&	3228	\\
V494	Sco	&	2453656.5500	&	0.4272919(4)	&	426.8(8)	&	455	&		&	-	&	0.055	&	-	&	1532	&	3185	\\
CD	Vel		&	2453748.7247	&	0.5735082(6)	&	66.27(5)	&		&	66.35	&	-	&	0.022	&	-	&	610	&	3294	\\
AF	Vel		&	2453747.8076	&	0.5274129(4)	&	58.62(7)	&	59	&	58.68	&	0.036	&	0.040	&	1.11	&	1028	&	3180	\\
V419	Vir	&	2452728.7022	&	0.5105245(8)	&	65.7(1)	&		&	65.69	&	0.041	&	0.042	&	1.02	&	413	&	3172	\\
AS	Vir *	&	2454631.5558	&	0.553412(1)	&	47.76(5)	&		&		&	-	&	0.055	&	-	&	420	&	3171	\\
SV	Vol		&	2454277.5415	&	0.6099111(8)	&	85.3(2)	&		&		&	0.031	&	-	&	-	&	1116	&	3300	\\
							\multicolumn{11}{c}{Stars	suspected of multiple	modulation}\\					
RU	Cet	&	2452117.8386	&	0.5862853(7)	&	98.0(2), 24.2(3)	&		&		&	0.046	&	0.047	&	1.02	&	430	&	3287	\\
SU	Col	&	2453396.6813	&	0.4873555(6)	&	65.22(7), 89.3(2)	&		&	65.6, 89.2	&	0.056, 0.043	&	0.039, 0.031	&	1.43, 1.39	&	924	&	3295	\\
IK	Hya	&	2453085.8054	&	0.650321(3)	&	71.81(5), 1403(21)	&	72	&	71.79	&	0.038, 0.45	&	0.059, 0.060	&	1.55, 1.33	&	633	&	3177	\\
RY	Oct	&	2452987.6056	&	0.5634473(7)	&	217(1), 26.88(5)	&		&		&	-	&	0.054, 0.036	&	-	&	649	&	3274	\\
AM	Vir	&	2453432.873	&	0.6150856(6)	&	49.61(5), 141.9(4)	&	49.8	&		&	-	&	0.023, 0.020	&	-	&	456	&	3163	\\
		\hline                  
\end{tabular}
\tablefoot{Star 1 means ASAS203420-2508.9, Star 2 means ASAS032438-2334.7 and Star 3 corresponds to ASAS062326+0005.8. See the text for more details 
						and column descriptions.}
\end{table*}

Modulation of some known Blazhko stars was not detectable in the ASAS sample. RZ~Cet, RV~Sex, V413~Oph, BB~Vir, SW Cru, V~Ind, and Z~Mic were only proposed to show modulation with no amplitude or period of modulation determined \citep{kovacs2005,samus2012,for2011,jurcsik1996}. Either they are stable or their modulation is of low amplitude. In V672 Aql we found a possible Blazhko period of 425.8~d, but the peak had $S/N<4$ and corresponding phase plot looked unconvincingly. Modulation of SS~Cnc, BR~Tau, RY~Com, and U~Cae was undetectable due to limited precision of ASAS data. VX~Her was studied by \citet{wunder1990}, who gave modulation period as 455.37~d and changes in $O-C$ with amplitude 0.013~d. No manifestation of modulation of VX Her was detected even in SW data. This could possibly be the consequence of ceasing Blazhko effect.
	
We should also mention AE PsA, which was a suspected Blazhko star with period 5.78~d (SF07). Frequency analysis exposed a peak with $f=2.0027$~c/d corresponding to a 5.78-day period, but this was probably a false peak, because it was close to a one-year alias of $f=2$~c/d. 

   \begin{figure}[htbp]
   \centering
   \includegraphics[width=8.5cm]{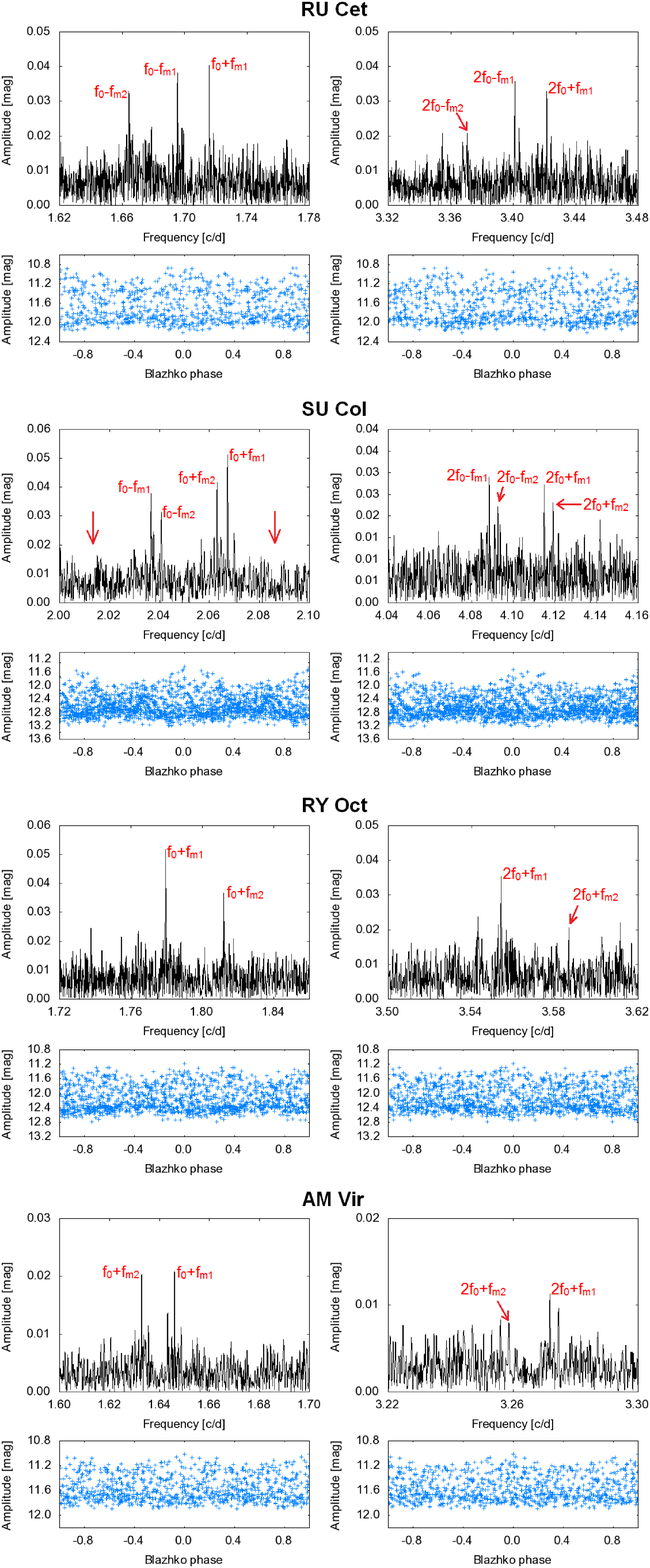}
   \caption{Four stars from our ASAS sample with double/irregular modulation. The first column depicts Fourier transform in the 						vicinity of $f_{0}$ after prewhitening with the main pulsation components and data phased with period corresponding 
   					to $f_{m1}$. There is the vicinity of $2f_{0}$ and phase plot according to the second modulation period (related to 
   					$f_{m2}$) in the second column.}%
   \label{multiASAS}
   \end{figure}
  	
\subsection{Stars suspected of multiple/irregular modulation and IK~Hya}

In the bottom part of Table \ref{table:1} there are six stars we suspect of some peculiarity in their modulations. All these stars showed additional peaks around $f_{0}$ probably related to an additional modulation component. This behavior was described, for instance, in XZ Cyg \citep{lacluyze2004}, UZ UMa \citep{sodor2006}, CZ Lac \citep{sodor2011}, or more recently in V445 Lyr \citep{guggenberger2012}. We also identified expected peaks around $2f_{0}$, which was a necessary condition to mark the stars as multiply modulated. Unfortunately, except for SU Col, these peaks had low amplitude under the confidence limit $S/N>4$. Prewhitened frequency spectra of these stars in the vicinity of $f_{0}$ and $2f_{0}$ with identification of the peaks are plotted in fig. 6. There are also phase-plots corresponding to suspected modulation periods in fig. \ref{multiASAS}.

SF07 identified SU Col as triple-modulated star. Peaks corresponding to periods 65 and 89~d (see fig. 6) were easily identified, while our analysis did not show any sign of the supposed 29.5-day period (the positions of peaks referring to this period are marked by an arrow in fig. 6). Residuals around additional peaks seem to be quite high, which can indicate that more peaks can be present but undetectable in the dataset. This, along with the absence of 29.5-day peak, could suggest that the Blazhko modulation of SU Col is irregular and changing.

\begin{figure}[htbp]
   \centering
   \includegraphics[width=8.5cm]{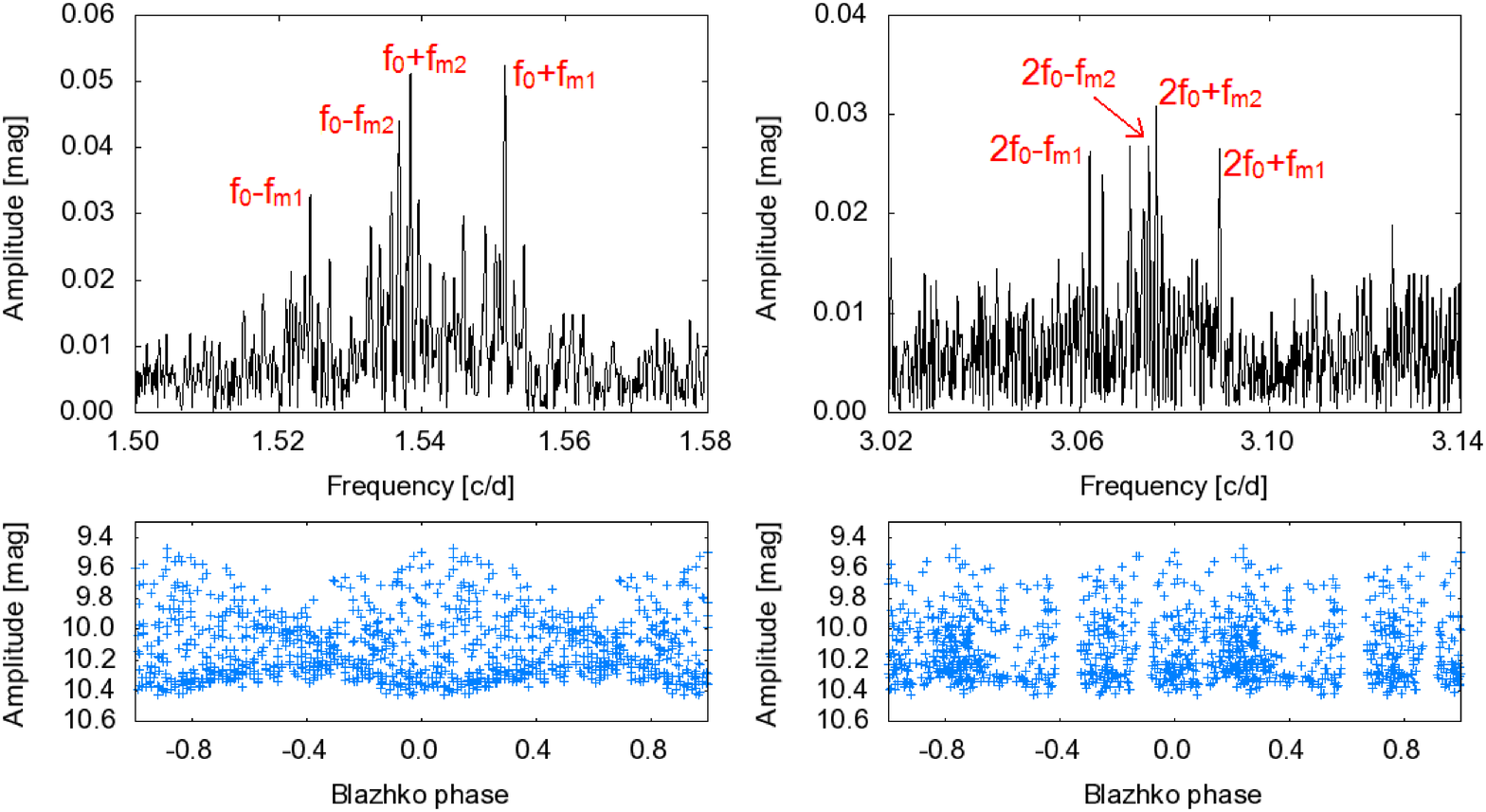}
   \caption{IK Hya. Vicinity of $f_{0}$, $2f_{0}$ and light curves constructed with periods 71.81~d (bottom left) and 
   			1403~d (bottom right). Frequency $f_{m1r}$ corresponds to 71.81~d, $f_{m1r}$ to 75.57~d , $f_{m2}$ to 1403~d.}%
   \label{IKHya}
   \end{figure}

A very interesting case of a possibly multiple/irregularly-modulated star is IK~Hya (fig. \ref{IKHya}). Its frequency spectra are somewhat confusing. There are two strong triplets ($f_{m1l}, f_{0}, f_{m1r}$ and $f_{m2l}, f_{0}, f_{m2r}$, where indexes $l$ and $r$ mean \textquoteleft left\textquoteright and \textquoteleft right\textquoteright, $f_{m1r}=f_{0}+f_{m1}$, $f_{m1l}=f_{0}-f_{m1}$ and similarly for $f_{m2l}$ and $f_{m2r}$). Frequency $f_{m1}$ corresponds to a 71.81-day known period, $f_{m2}$ holds for the newly determined period of 1403~d. The spacing of the closer triplet was almost identical; i.e., $\Delta f=\left( f_{m2r}-f_{0}\right)-\left( f_{0}-f_{m2l}\right)=4\cdot10^{-5}$, while the offsets of wider peaks from $f_{0}$ were identical only roughly ($\Delta f=6.9 \cdot 10^{-4}$). This discrepancy resulted in an inequality of the first modulation period based on the left peak (75.57~d) and the right peak (71.81~d). Thus, these peaks might indicate two independent periods or limiting values of changing modulation period.

This is supported by the identification of symmetric counterparts of $f_{m1l}$ and $f_{m1r}$ with respect to $f_{0}$, which 
we found in frequency spectra, but which had amplitudes close to $S/N=4$. The difference between the right-hand-side peaks $f_{m1r}$ and $f_{m2r}$ nearly equals $f_{0}-f_{m1l}$, in other words $75.57^{-1}\cong71.81^{-1}+1403^{-1}$. This means that we actually observed one narrow triplet $f_{m2l}, f_{0}, f_{m2r}$ and two doublets (one on each side of $f_{0}$) with spacing corresponding to 1403-d period. This could indicate a 1403-day long-term cycle, during which the Blazhko and pulsation characteristics are changing similarly to the four-year cycle of RR~Lyr \citep{detre1973,stellingwerf2013}. A very interesting and possibly important thing is that the long period (1403~d) is very close to half of the beating period of 75.57~d and 71.81~d periods, which is 2886~days.

\begin{table*}[htbp]
 	\caption{New Blazhko stars from our ASAS sample}               
	\label{table:2}      
	\centering      
	\begin{tabular}{l c l l c c c c c}     
	\hline       
\multicolumn{1}{l}{Star}&	Epoch HJD &	\multicolumn{1}{c}{$P_{\mathrm{puls}}$ [d]}	& \multicolumn{1}{c}{$P_{\mathrm{BL}}$ [d]}		
&	$A_{-}$ [mag]&	$A_{+}$ [mag]	&$A_{+}/A_{-}$&	$n$	&	$TS$ [d] \\
\hline       
EL	Aps	&	2451870.4229	&	0.5797216(5)	&	25.96(2)&					-		&	0.019	&	-	&	849	&	3275	\\
HH	Aqr	&	2451889.5259	&	0.574434(1)	&	3360(270)	&					-		&	0.057	&	-	&	395	&	3274	\\
LS	Boo	&	2453103.7768	&	0.552704(2)	&	42.28(9)	&					-		&	-	&	-	&	359	&	2371	\\
UY	Boo	&	2454289.5426	&	0.650902(2)	&	171.8(2)	&					-		&	0.038	&	-	&	312	&	2907	\\
V595 Cen&	2453835.6569	&	0.691035(1)	&	107.5(6)	&					-		&	0.025	&	-	&	656	&	3188	\\
RW	Hyi	&	2454150.5160	&	0.555791(1)	&	135.2(6)	&					0.048	&	0.036	&	0.75	&	654	&	3296	\\
AO	Lep	&	2453767.6882	&	0.5600866(7)	&	93.0(1)	&					0.048	&	0.044	&	0.92	&	570	&	3296	\\
XZ	Mic	&	2452757.8704	&	0.4491584(4)	&	85.7(3)	&					-		&	-	&	-	&	595	&	3238	\\
V784 Oph&	2452775.7903	&	0.603359(1)	&	25.61(2)	&					-		&	0.044	&	-	&	352	&	2884	\\
ASAS173154-1653.1	&	2453864.7477	&	0.602590(1)	&	52.41(8)	&	-	&	0.014	&	-	&	453	&	3145	\\
CY	Peg	&	2453337.5595	&	0.647939(2)	&	464(5)	&					0.071	&	-	&	-	&	238	&	2349	\\
RU	Scl	&	2452141.8678	&	0.4933549(2)	&	23.91(2)	&					-	&	0.010	&	-	&	338	&	3262	\\
AE	Scl	&	2452213.5875	&	0.5501123(9)	&	46.2(2)	&					-		&	0.046	&	-	&	497	&	3243	\\
V	Sex		&	2454633.4803	&	0.487372(1)	&	118.5(4)	&					-		&	0.061	&	-	&	317	&	2570	\\
SS	Tau	&	2453046.5674	&	0.3699068(7)	&	123.3(4)	&					-	&	0.057	&	-	&	320	&	3220	\\
HH	Tel	&	2451963.8028	&	0.4820925(5)	&	143(2)	&					-		&	-	&	-	&	518	&	3166	\\
FS	Vel	&	2454618.5425	&	0.4757576(5)	&	59.29(7)	&					-	&	0.040	&	-	&	618	&	3300	\\
UV	Vir	&	2451887.0160	&	0.5870842(7)	&	103.9(3)	&					-	&	0.027	&	-	&	365	&	3163	\\
AF	Vir	&	2454678.5669	&	0.483747(1)	&	42.15(5)	&					-		&	0.041	&	-	&	340	&	2915	\\		\hline                  
\end{tabular}
\end{table*}	
				
In SW data only one slightly asymmetric triplet (Blazhko effect with period 73.2~d) was identified. This period would support the theory of changing Blazhko effect of IK Hya rather than compound modulation with three periods. With ASAS and SW data, it is impossible to perform a more detailed analysis. Therefore, IK~Hya is an object that deserves to be investigated much more closely with larger telescopes than for automated surveys.

	\subsection{New Blazhko stars}

There are 19 new Blazhko stars in Table \ref{table:2}, which were not identified by WS05 or by SF07. Only two of the newly revealed Blazhko variables showed a triplet structure, which is probably a consequence of weak Blazhko effect manifestation. V784 Oph is actually not a new Blazhko star, because we discovered the modulation at the same time as \citet{deponthiere2013}, who identified its multiple modulation (our analysis showed only simple modulation).

   \begin{figure}[htbp]
   \centering
   \includegraphics[width=8.3cm]{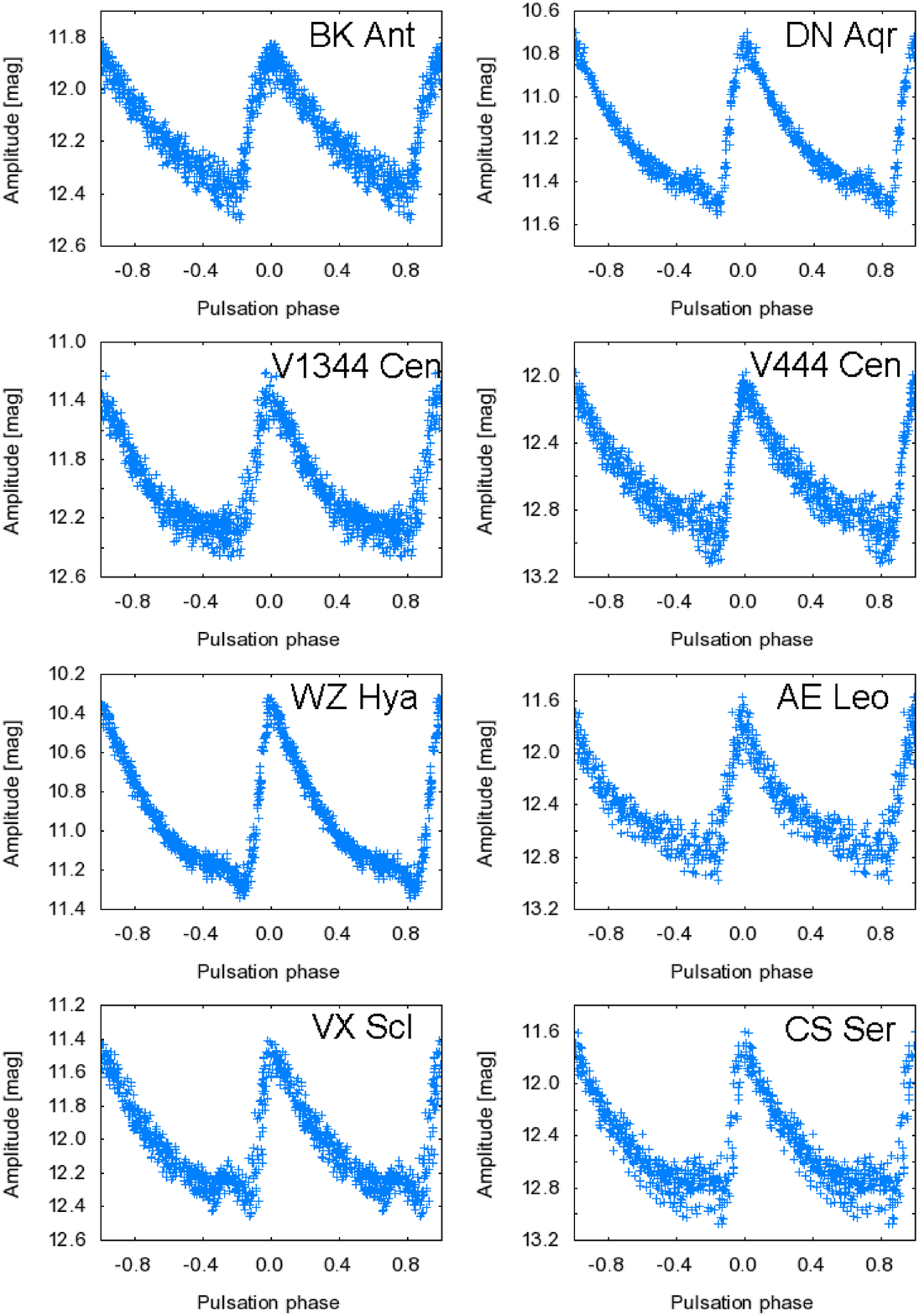}
   \caption{Possible Blazhko stars. Data are phased with respect to ephemerides given in Table \ref{table:3}.}%
   \label{posBlazhko}
   \end{figure}
		
\begin{figure*}
   \centering
   \includegraphics[width=18cm]{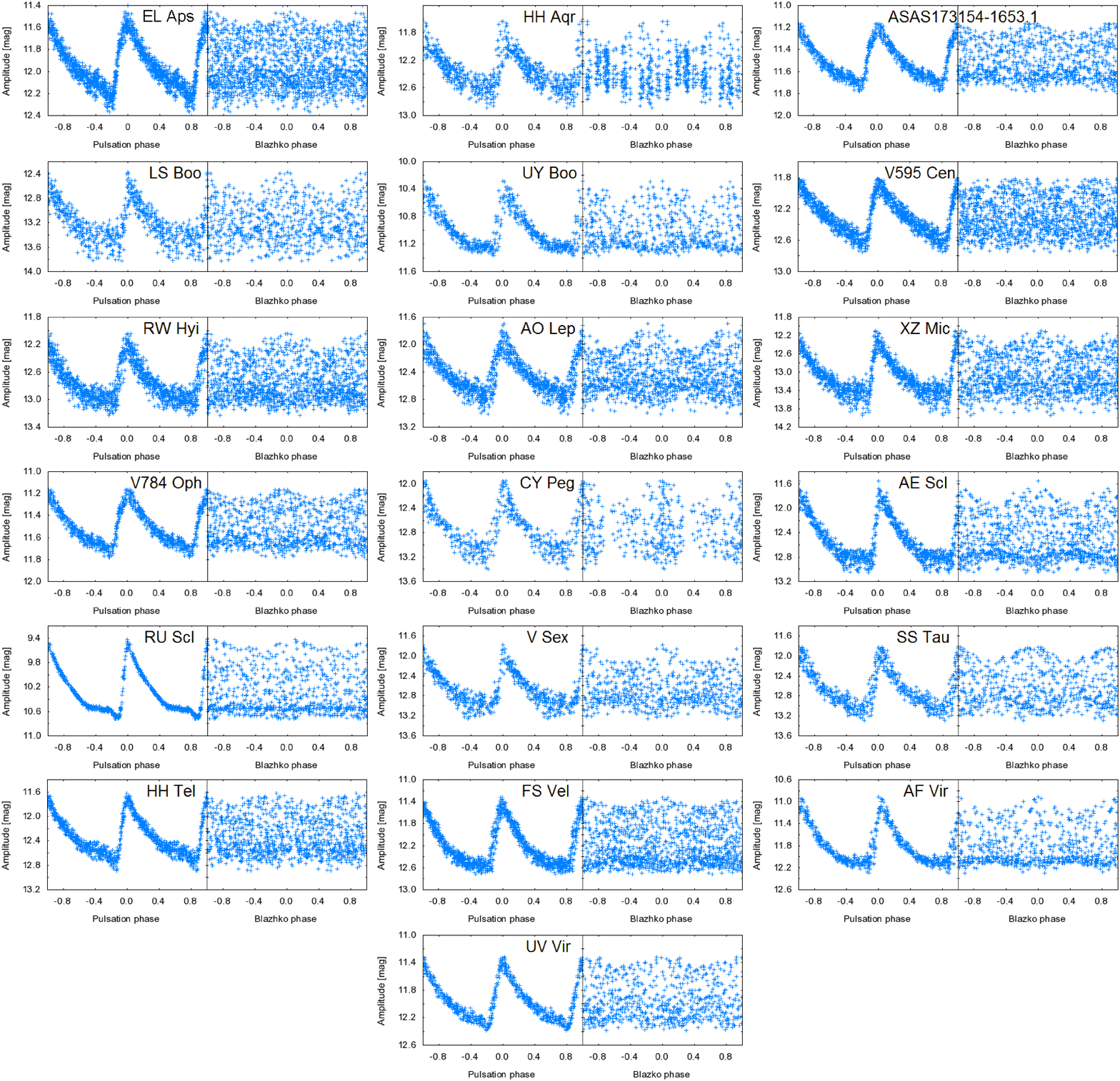}
   \caption{New Blazhko stars from ASAS database. Data are phased with respect to ephemerides given in tab. 2.}%
   \label{newBlASAS}
   \end{figure*}

In the case of LS Boo, XZ Mic, and HH Tel, their modulation periods are based on the side peaks with $S/N$ slightly lower than 4, but their Blazhko phased light curves look good (fig. \ref{newBlASAS}). Comparing the Blazhko periods of HH~Tel and of XZ Mic with values coming from the SW data (Table \ref{table:5}) we see that they are real and nearly have the same value. This raises the question, if the significance condition $S/N>4$ should not be lowered\footnote{as done for example in \citep{kolenberg2006}, who used criterion $S/N=3.5$ for modulation peaks}. HH~Aqr have a somewhat uncertain Blazko period, because it is longer than the time span of the data. 

Frequency analysis of UY Boo showed, except for the modulation side peak, also another additional peak, which corresponds to the period 1976~d. This is probably the consequence of its peculiar period changes \citep{leborgne2007}.

\def\arraystretch{1}
\setlength\tabcolsep{1ex}
	
\begin{table}[htbp]
 	\caption{Stars suspected of the Blazhko effect}               
	\label{table:3}      
	\centering  
    
	\begin{tabular}{M M M M M}     
	\hline  \multicolumn{1}{l}{Star}&	\multicolumn{1}{c}{Epoch HJD} &	\multicolumn{1}{c}{$P_{\mathrm{puls}}$ [d]}	&	\multicolumn{1}{c}{$n$}	
	&	\multicolumn{1}{c}{$TS$ [d]} \\
	\hline
BK	Ant	&	2454911.6407	&	0.5165704(6)													&	602	&	3169 \\
DN	Aqr	&	2452471.8336	&	0.6337573(4)													&	400	&	3219 \\
V444	Cen	&	2452144.6919	&	0.5141328(6)													&	537	&	3194 \\
V1344	Cen	&	2453890.5987	&	0.4182563(4)													&	594	&	3187 \\
WZ	Hya	&	2454838.7787	&	0.5377193(3)													&	628	&	3300 \\
AE	Leo	&	2452630.7110	&	0.626673(2)													&	421	&	2400 \\
VX	Scl	&	2451916.0250	&	0.637066(1)													&	552	&	3295 \\
CS	Ser	&	2453896.6210	&	0.5267982(8)													&	472	&	3168 \\ \hline
\end{tabular}
\end{table}

The Blazhko effect of stars in Table \ref{table:3} is somewhat ambiguous. These eight stars are candidate Blazhko variables. They show 
scattering around maxima (fig. 8), but not necessarily due to modulation -- it can simply be the consequence of the scattering. There were peaks with $S/N<4$ identified in the Fourier spectrum of CS Ser, which could be caused by the Blazhko effect with period 118.3~d. Similar to the aforementioned cases of HH Tel and XZ Mic with the same characteristics, these peaks probably are real manifestations of the modulation. 

\def\arraystretch{1.2}
\setlength\tabcolsep{1ex}
	\begin{table*}
 	\caption{Known Blazhko stars from our SW sample.}               
	\label{table:4}      
	\centering  
    
	\begin{tabular}{M M M M M M M M M M M}
	\hline       

Star &	Epoch HJD &	$P_{\mathrm{puls}}$ [d]	& $P^{\mathrm{F}}_{\mathrm{BL}}$ [d]		& $P^{\mathrm{M}}_{\mathrm{BL}}$ &	
$A_{-}$ [mag]&	$A_{+}$ [mag]	&$A_{+}/A_{-}$&	n	&	TS [d] & $P^{\mathrm{Lit}}_{\mathrm{BL}}$ \\
\hline
DR And	&	2454334.5942	&	0.563112(6)		&	58.11(2)	&	58.2(3)	&	0.052	&	0.066	&	1.27	&	4536	&	705		&	57.5$^{(1)}$		\\
U Cae*	&	2453996.5377	&	0.41978515(5)	&	22.77(2)	&	22.7(5)	&	0.025	&	-			&	-			&	5252	&	491		&	22.8$^{(2)}$		\\
AH Cam	&	2454362.6540	&	0.368721(4)		&	10.772(6)	&	10.65(6)&	0.034	&	0.075	&	2.21	&	1783	&	1227	&	10.83$^{(2)}$		\\
SS Cnc*	&	2454167.5509	&	0.367341(1)		&	5.419(9)	&	5.42(3)	&	0.012	&	-			&	-			&	3554	&	116		&	5.309$^{(3)}$		\\
Z Cvn		&	2456078.3420	&	0.653684(9)		&	22.48(2)	&	-				&	0.042	&	0.044	&	1.04	&	2086	&	121		&	22.98$^{(2)}$		\\
SS CVn	&	2453140.6137	&	0.4785195(2)	&	94.19(2)	&	94.26(9)&	0.062	&	0.099	&	1.54	&	8811	&	1476	&	93.72$^{(2)}$		\\
RY Col*	&	2454423.3675	&	0.4788302(2)	&	82.35(2)	&	82.4(2)	&	0.027	&	0.074	&	2.74	&	9478	&	516		&	82.08$^{(2)}$		\\
DM Cyg	&	2454344.4237	&	0.4198639(8)	&	10.38(1)	&	10.7(1)	&	-			&	0.009	&	-			&	1940	&	473		&	10.57$^{(4)}$		\\
XZ Dra	&	2454686.4674	&	0.4765587(5)	&	-					&	75(1)		&	-			&	-			&	-			&	5807	&	83		&	73--77$^{(5)}$	\\
SS For*	&	2453997.4793	&	0.4954351(5)	&	34.73(1)	&	34.66(7)&	0.03	&	0.059	&	1.97	&	8684	&	654		&	34.94$^{(6)}$		\\
RT Gru*	&	2454246.6550	&	0.5121845(6)	&	86.7(2)		&	87.8(3)	&	0.025	&	0.037	&	1.48	&	10238	&	537		&	87$^{(7)}$			\\
SV Hya*	&	2454528.4773	&	0.478546(2)		&	62.8(3)		&	63(2)		&	-			&	0.025	&	-			&	5374	&	131		&	63.29$^{(8)}$		\\
IK Hya*	&	2454155.5952	&	0.650247(4)		&	73.18(3)	&	73.2(3)	&	0.032	&	0.061	&	1.91	&	12856	&	754		&	67.5$^{(2)}$		\\
V559 Hya*&	2454199.5359	&	0.4479571(2)	&	26.454(2)	&	26.47(2)&	0.093	&	0.12	&	1.29	&	10055	&	754		&	26.6$^{(7)}$		\\
FU Lup	&	2454594.3086	&	0.3821526(3)	&	42.81(1)	&	42.74(9)&	0.034	&	0.107	&	3.15	&	8300	&	752		&	42.49$^{(8)}$		\\
V1645 Sgr*&	2454307.315	&	0.552955(3)		&	-					&	-				&	-			&	-			&	-			&	11310	&	754		&	1331.74$^{(8)}$	\\
BR Tau*	&	2454143.386		&	0.390600(1)		&	-					&	18.14		&	-			&	-			&	-			&	2063	&	139		&	19.3$^{(9)}$		\\
CD Vel*	&	2454151.3312	&	0.5735080(7)	&	66.34(4)	&	66.38(26)&	0.016	&	0.023	&	1.44	&	14501	&	735	&	66.35$^{(8)}$		\\
	\multicolumn{11}{c}{Stars	suspected of multiple	modulation}		\\
RS Boo	&	2453132.5463	&	0.3773210(2)	&	41.3(2), 62.5(2)	&	64(3)	&	0.006, 0.006	&	0.010, 0.007	&	1.66, 1.16	&	3737	
				&	133	&	532.48$^{(2)}$	\\
RW Cnc	&	2454118.5843	&	0.547191(6)		&	29.14(8), 21.9(1)	&	30.0(3), 21.9(5)	&	0.021	&	0.012,0.014	&	0.57	&	3513	
				&	129	&	87$^{(10)}$	\\
RX Col*	&	2454106.4342	&	0.593743(1)		&	134.3(1), 79.7(2)	&	135.7(7), 81(2)	&	0.014	&	0.050, 0.011	&	3.57	&	11003	
				&	677	&	134.77$^{(2)}$	\\
RW Dra	&	2454655.5652	&	0.4429239(4)	&	41.42(1), 72.6(2)	&	41.49(5), 72.3(3)	&	0.055, 0.018	&	0.090, 0.013	&	1.64, 0.72	
				&	17386	&	492	&	41.42$^{(2)}$	\\
\hline                  
\end{tabular}

\tablebib{(1)~\citet{lee2001}; (2)~\citet{leborgne2012}; (3)~\citet{jurcsik2006}; (4)~\citet{jurcsik2009b};
					 (5)~\citet{jurcsik2002}; (6)~\citet{kolenberg2009}; (7)~WS05; (8)~SF07; (9)~\citet{jurcsik2009a}; 
					 (10)~\citet{smith1995}.}

\end{table*}

		
\section{Blazhko stars from SW}	
	   
	\subsection{Known Blazhko stars}	
		
After a pre-selection of 243 RR Lyraes from SW data we obtained 106 stars for further analysis. Twenty-eight objects were proposed to show the Blazhko effect. For six of them (SW~And, OV~And, V~Ind, U~Tri, RY Com, and AR Per), no significant indication of the modulation was found. In RY Com and AR Per, this was due to limited precision of used data, because they are modulated stars \citep{jurcsik2009a,vsx}.

Twenty-two remaining known Blazhko stars are listed in Table \ref{table:4}. The fifth column gives the modulation period based on frequency 
analysis ($P^{\mathrm{F}}_{\mathrm{BL}}$), and the sixth ($P^{\mathrm{M}}_{\mathrm{BL}}$) gives the Blazhko period based on the $Mag_{\mathrm{max}}$ vs. $T_{\mathrm{max}}$ analysis. The last column provides modulation periods known from literature. Other columns correspond with those in Table \ref{table:1}. An asterisk in the first column (if present) indicates that the data were also available in the ASAS survey. The times ($T_{\mathrm{max}}$) and magnitudes $Mag_{\mathrm{max}}$ of maxima (Table \ref{table:5}) can be found in electronic form as an on-line material at the CDS portal.
		
\begin{table}[htbp]
 	\caption{Times and magnitudes of maximum light of 35 stars from SW survey, which is available in electronic form.}  
	\label{table:5}      
	\centering     
	\begin{tabular}{c c c c c}    
	\hline  Star & HJD $T_{\mathrm{Max}}$ & err$_{\mathrm{HJD}}$ & Max [mag] & err$_{\mathrm{Max}}$ \\
	\hline
 \\
DR And & 2454325.5842&4&11.972&4 \\
			 & 2454330.6559&9&11.832&12 \\ 
... & ... & ... & ... & ...\\\hline
\end{tabular}
\end{table}
					
Frequency spectra of almost all SW stars were affected by artificial effects, such as small shifts in the data from different nights or by gaps between observations, which resulted in additional peaks near $f_{0}$ with no clear interpretation. Peaks corresponding to the time span of the data occurred frequently.

In XZ Dra and BR Tau, the Blazhko effect was not detected in their frequency spectra, but it was noted in the change of maximum magnitude. V1645 surely showed modulation, but SW data were in this case insufficient for determining the Blazhko period.

\subsection{SW Stars suspected of multiple/irregular modulation}
 
   \begin{figure*}[htbp]
   \centering
   \includegraphics[width=18cm]{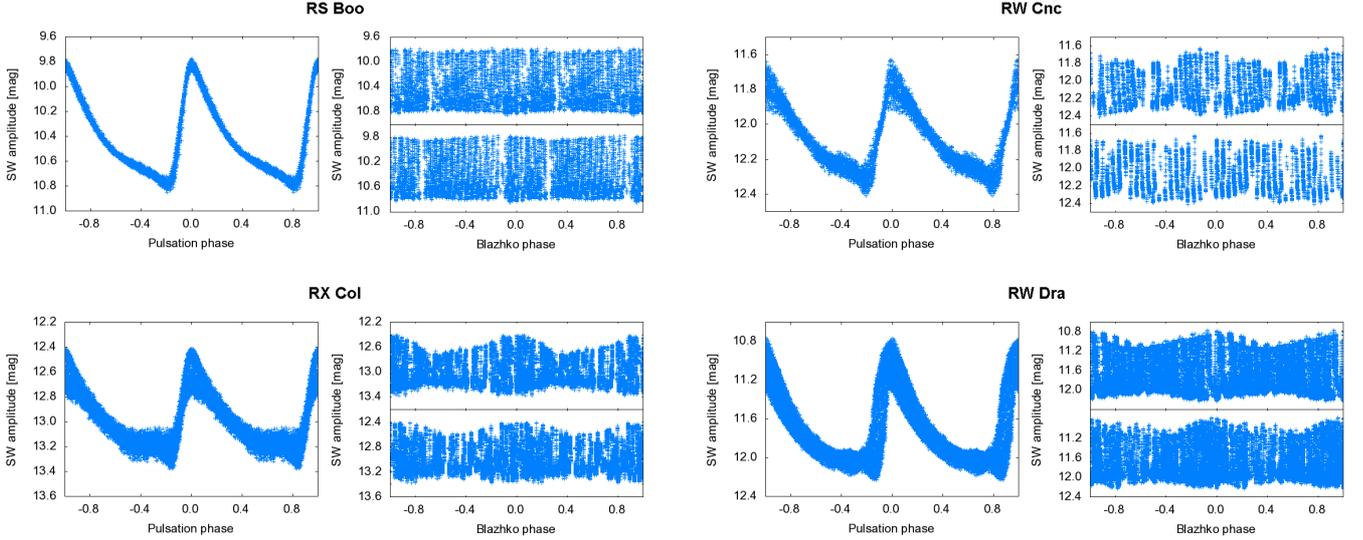}
   \caption{Known Blazhko stars with more than one modulation period. In the left panel of each star the data are folded with 
   					respect to the basic pulsation period. On the left, panels are data-phased with modulation periods.}%
   \label{multiSW}
   \end{figure*}

Four known Blazhko variables in our SW sample were identified as stars with multiple modulation (the bottom part of Table \ref{table:4}, fig. \ref{multiSW}). RS~Boo, RW~Cnc and RW~Dra were suspected of such behaviour in the past; RX~Col is a newly discovered as multiple modulated star (it was known as single-modulated star). 
		
The Blazhko period of $\sim$533~d of RS Boo has been known for a long time \citep{oosterhoff1946}, but there have been some indications of additional periodicity. \citet{kanyo1980} reported another period in the range of 58-62~days, but \citet{nagy1998} did not observe it. There was no indication of a 533~d period in SW data -- probably due to the time span of only 133~d. During the analysis we revealed a period of 62.5~d, which is close to the period noted by \citet{kanyo1980}. Except for this period we also noted obvious indications of a period of 41.3~d. 
		
\citet{balasz1950} reported RW Cnc to be a double-modulated star with modulation periods 29.95 and 91.1~d. The first modulation 
period was determined as 29.14~d (30.0~d from maxima analysis), and the second period 21.9~d from both frequency and maxima analysis. 
No indications of a 91-day cycle were found. The 87-day period given by \citet{smith1995} was also not observed. 
		
RW Dra was found to be multiple-modulated by \citet{balasz1952}. A Blazhko period of 41.61~d was accompanied by periods of 124~d 
and two other periods with lengths of years. We noted the Blazhko period of 41.42~d and only one secondary modulation period with a length of 72.6~d. 
	
These four stars have a very interesting feature -- the modulation periods are very close to a resonance with small integers: 
$\left(P_{m2}/P_{m1}\right)_{\mathrm{RS~Boo}}=1.513\cong$~3:2; $\left(P_{m1}/P_{m2}\right)_{\mathrm{RW~Cnc}}=1.331\cong$~4:3; 
$\left(P_{m1}/P_{m2}\right)_{\mathrm{RX~Col}}=1.685\cong$~5:3 and $\left(P_{m2}/P_{m1}\right)_{\mathrm{RW~Dra}}=1.753\cong$~7:4. 
In the ASAS stars only the modulation periods of SU Col approximate the resonance 4:3 ($P_{m2}/P_{m1}=1.369$). This seems very odd. Either the resonances are some strange consequence of SW data or the stars behave similarly to CZ Lac, where resonances 
of the modulation periods were also observed \citep{sodor2011}.

\def\arraystretch{1.2}
	\begin{table*}
 	\caption{New Blazhko stars from our SW sample.}               
	\label{table:6}      
	\centering     
	\begin{tabular}{M M M M M M M M M M}     
	\hline 
Star &	Epoch HJD &	$P_{\mathrm{puls}}$ [d]	& $P^{\mathrm{F}}_{\mathrm{BL}}$ [d]		& $P^{\mathrm{M}}_{\mathrm{BL}}$ &	
$A_{-}$ [mag]&	$A_{+}$ [mag]	&$A_{+}/A_{-}$&	n	&	TS [d]  \\ 
	\hline 
XX Boo		&	2454267.5088	&	0.5813995(3)&	105.9(1)	&	105.6(3)	&	0.012	&	0.015	&	1.25	&	8707	&	1150\\
V595 Cen*	&	2454239.2256	&	0.6910344(5)&	107.41(9)	&	107.6(2)	&	0.004	&	0.019	&	4.75	&	15709	&	754	\\
GY Her		&	2454619.5950	&	0.524244(6)	&	49.6(2)		&	49.1(9)		&	0.049	&	0.053	&	1.08	&	8065	&	108	\\
XZ Mic*		&	2453964.2968	&	0.4491569(6)&	85.8(2)		&	85.7(5)		&	0.034	&	0.035	&	1.03	&	8591	&	537	\\
TU Per		&	2454438.5346	&	0.6070469(4)&	59.0(3)		&	-					&	0.021	&	0.015	&	0.71	&	2383	&	108	\\
RU Scl*		&	2454364.4311	&	0.493368(1)	&	23.6(1)		&	24.0(3)		&	0.003	&	0.008	&	2.67	&	3725	&	103	\\
V4424 Sgr	&	2453881.6352	&	0.4245024(2)&	29.28(4)	&	27.62(8)	&	0.005	&	0.004	&	0.80	&	8778	&	528	\\
HH Tel*		&	2454347.2643	&	0.482056(5)	&	-					&	133(9)		&	-			&	-			&	-			&	2761	&	93	\\
FS Vel*		&	2454201.2979	&	0.4757597(2)&	59.48(4)	&	59.1(2)		&	0.020	&	0.026	&	1.30	&	9677	&	713	\\
						\multicolumn{10}{c}{Stars	suspected of multiple/irregular modulation}												\\
RZ CVn		&	2454263.4059	&	0.567418(-)&	33.35-33.26	& 29.6-33.36	&	0.007	&	0.017	
					&	2.43	&	7148	&	1136	\\
AG Her		& 2454669.4418	& 0.6494465(3)& 79.69(7), 47.93(7) & 79.58(7),47.96(2) & 0.010, 0.014 & 0.030, 0.020
					& 3, 1.43 & 22362 & 1552 \\					
AE Scl*		&	2453920.5919		&	0.550116(2)	&	46.07(2), 104.7(2)	&	46.04(4)	&	0.019, 0.019	&	0.044, 0.015	
					&	2.32, 0.79	&	8179	&	563	\\
\hline                  
\end{tabular}
\end{table*}

	\subsection{New Blazhko stars from SW}

Nine new, previously unknown single-modulated Blazhko stars, were identified in our SW sample (Table \ref{table:6}, fig. \ref{newSW}). For six of these stars (marked by an asterisk) ASAS data were also available. Three stars showed double modulation or some peculiarity in their Blazhko effect. 

\begin{figure}[htbp]
   \centering
   \includegraphics[width=8.2cm]{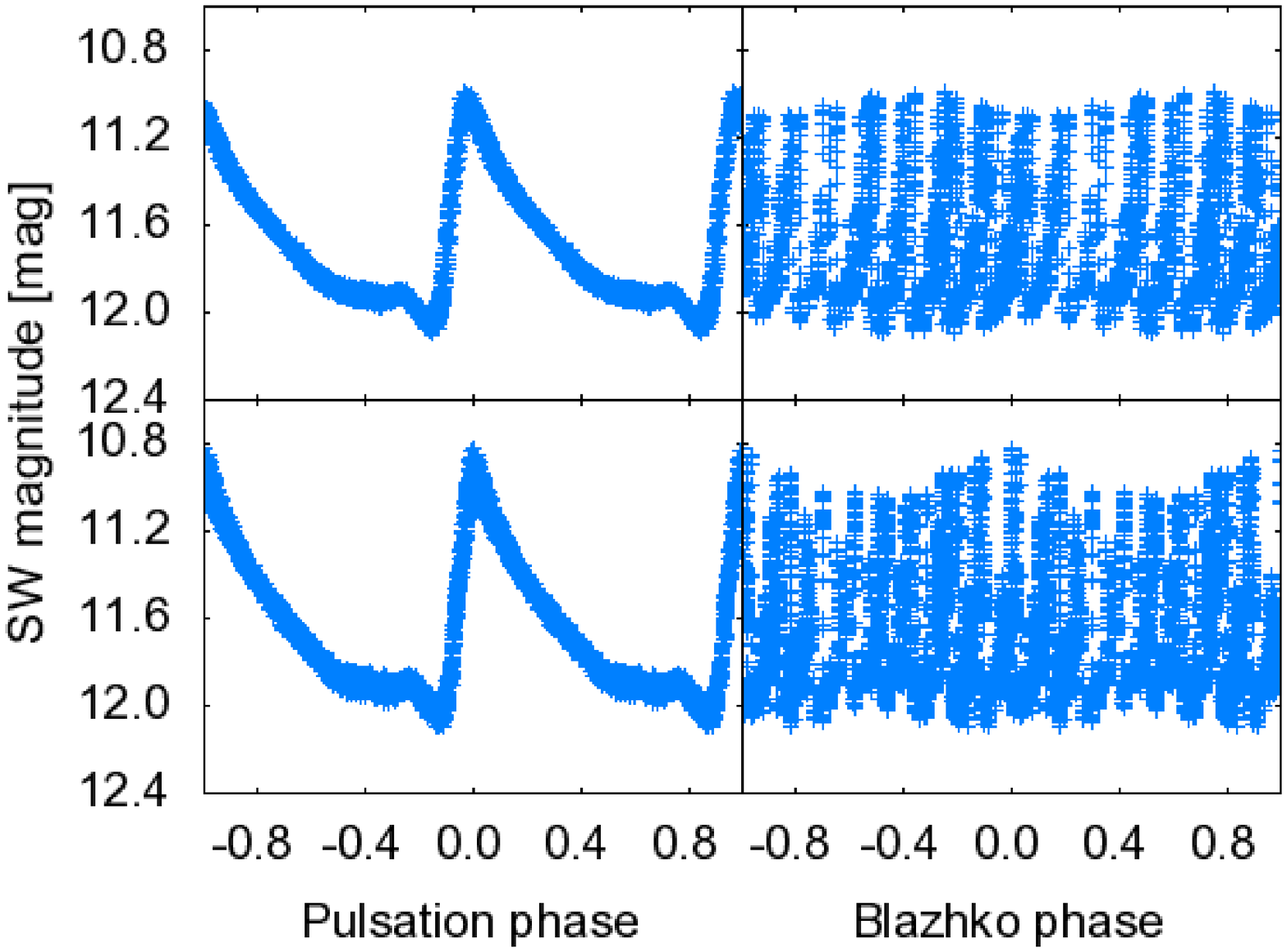}
   \caption{RZ CVn in different seasons. Change in the light variations, as well as in the amplitude of the modulation is 			
   					noticeable.}%
   \label{RZCVn}
   \end{figure}

\begin{figure*}[htbp]
   \centering
   \includegraphics[width=18cm]{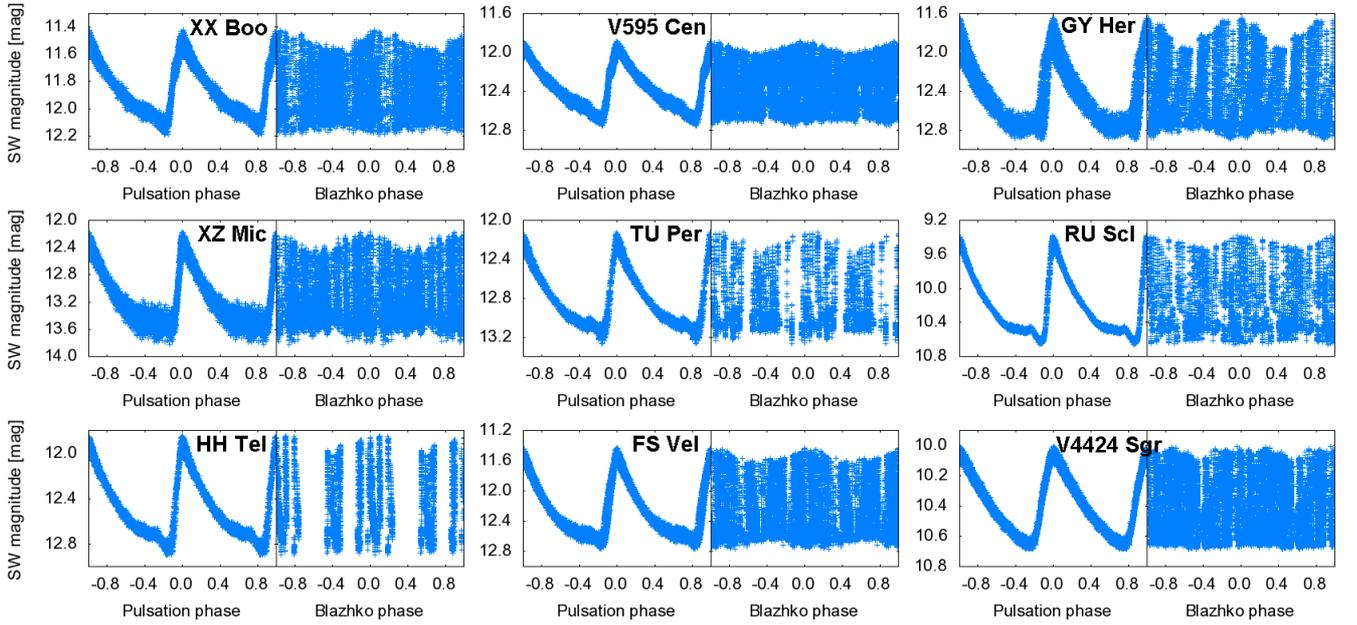}
   \caption{New Blazhko stars from SW database. Light curves are constructed according to ephemerides listed in Table \ref{table:6}.}%
   \label{newSW}
   \end{figure*}

RZ Cvn (fig. \ref{RZCVn}) seemed to change its Blazhko period. In the first part of the used data (between HJD 2453130 and 2453249), the Blazhko period was 33.35~d (29.6~d on the basis of maxima analysis), the amplitude of the modulation was 0.104~mag (peak to peak) and $A_{1}=0.339$~mag, while between HJD 2454135 and 2454266, the Blazhko cycle lasted 33.26~d (33.4~d based on maxima analysis). The amplitude of the modulation was almost twice as big as in the first part (0.192~mag) and $A_{1}=0.363$~mag. The basic pulsation period also changed slightly (0.567411 vs. 0.567418~d). Similar behaviour is known in, say, XZ~Dra \citep{jurcsik2002} or RV~UMa \citep{hurta2008}. We point out that in the case of RZ CVn, the Blazhko periods determined on the basis of maxima analysis offer better confidence.
		
The second multiple-modulated star, AG Her, shows possible resonances of its modulation periods $\left(P_{m1}/P_{m2}\right)_{\mathrm{AG~Her}}=1.659\cong$~5:3 similar to RX~Col, and another three stars discussed in previous section. AE Scl was identified as single-modulated in the ASAS data set.

		\begin{figure}[htbp]
   		\centering
   		\includegraphics[width=8cm]{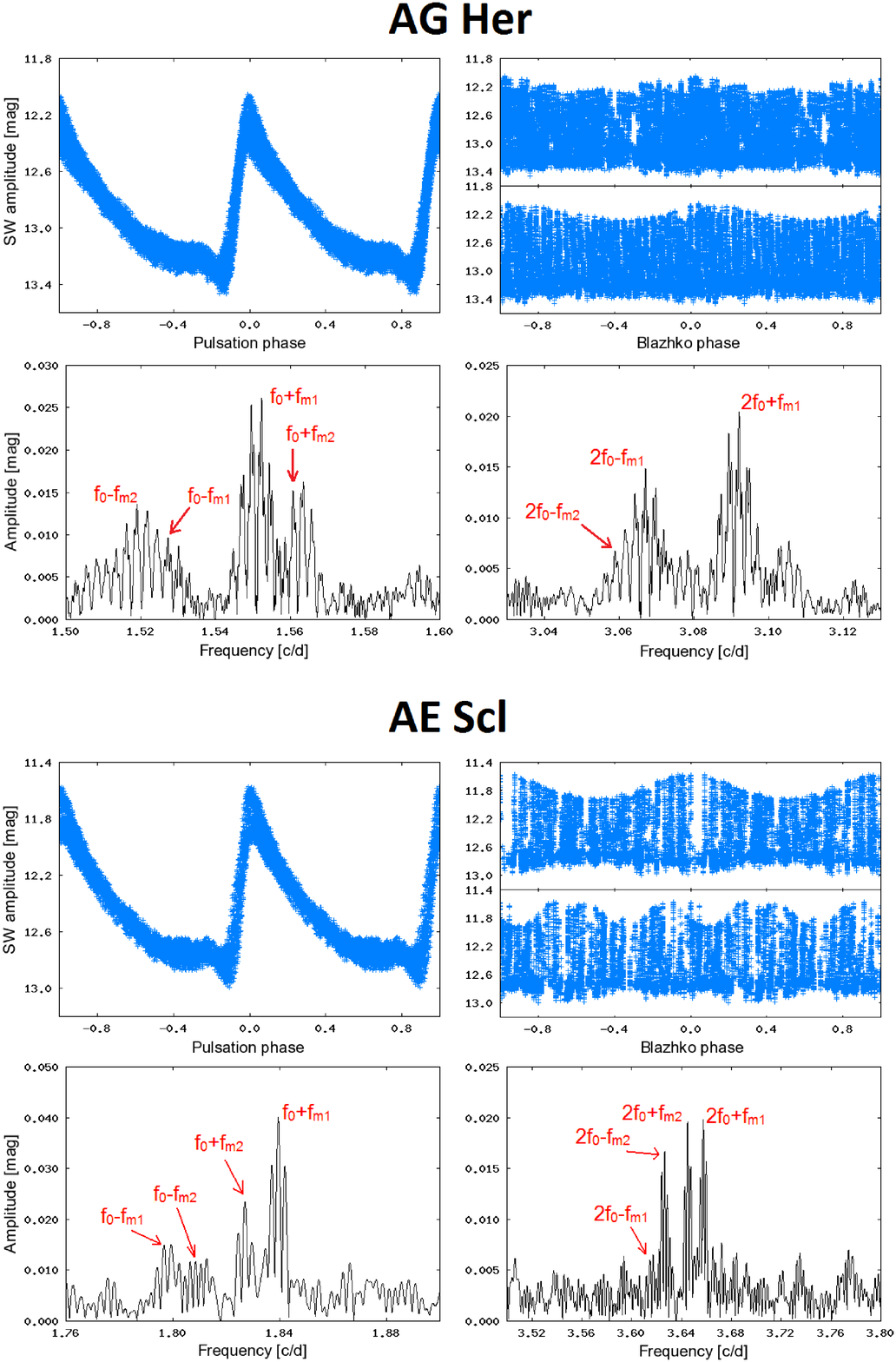}
   		\caption{Data of AG Her and AE Scl folded with elements given in Table \ref{table:6} and with their frequency spectra in the vicinity 
   						of $f_{0}$ and $2f_{0}$. In Blazhko phased data of AE Scl it is apparent that the second modulation period should instead
   						be 104.7/2 than 104.7~d, but the frequency spectra indicated period 104.7~d.}%
   		\label{AGHer}
   	\end{figure}
   	
	\section{Statistics of the Blazhko variables in the sample}

From the sample containing 321 RRab type stars, one hundred objects were identified as Blazhko variables. This means that the incidence rate of Blazhko stars among RRab sample stars was 31~\%. If we include certain Blazhko variables, whose modulation was not detectable (5), this percentage would grow almost to 33~\%. If Blazhko candidates (8) are also taken into account, this rate would be 35~\%. Finally, if all suspected stars are included (18 instead of 5), the percentage would be 39~\%. In addition, it is likely for this number to be slightly higher, in fact, because the modulation of a number of stars with low modulation amplitude might remained uncovered. 
	
Compared to incidences 5.1~\% (SF07) and 4.3~\% \citep{wils2006}, our percentage is significantly higher and roughly approximates the percentage based on precise measurements -- 48~\% \citep{benko2010} based on Kepler data, or 47~\% \citep{jurcsik2009a}, who utilized precise ground-base observations.
		
Our incidence is also higher than in LMC \citep[20~\%, ][]{soszynski2009} or than in SMC \citep[22~\%, ][]{soszynski2010}. In the Galactic bulge this rate is about 25~\% \citep{mizerski2003}. It is worth mentioning that these numbers almost doubled during the past few years owing to more sensitive analysis techniques and an increase in the time span of the data. It is likely that these numbers will also increase slightly in future works.
	
About 60~\% of confirmed Blazhko variables were of BL2 type\footnote{Peaks related to the modulation distributed on both sides of basic pulsation components \citep{alcock2003}}. Considering ASAS and SW surveys separately, this rate was 48~\% (ASAS) and 76~\% (SW). Several stars in the SW sample turned out to be of BL2, while in ASAS they were BL1 type. This finding clearly shows that classification of the stars based on their side-peak distribution strongly depends on data quality and should not be used when analysing low-quality data, which is known \citep{alcock2003}. In addition, ultra-precise space measurements showed that almost every modulated star\footnote{V2178 and V354 Lyr are possibly of BL1 type \citep{nemec2013}.} has peaks on both sides. Our recommendation is also supported by the fact that the percentage of stars, which show $A_{+}/A_{-}>1$, was about 89~\% in our ASAS sample contrary to 51\% given by SF07.	
	
\begin{figure}[htbp]
   \centering
   \includegraphics[width=9cm]{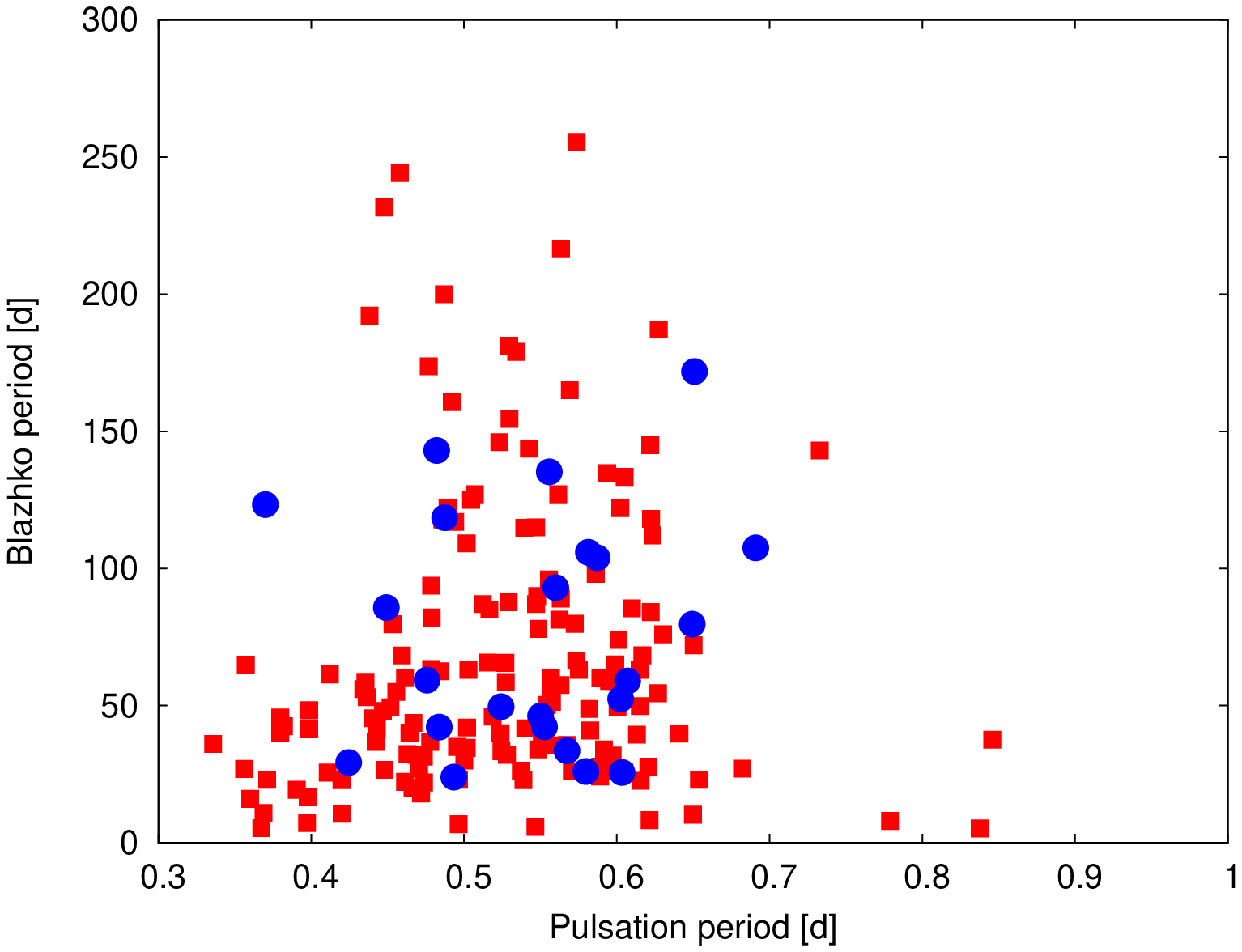}
   \caption{$P_{\mathrm{BL}}$ vs. $P_{\mathrm{puls}}$ diagram of known RRab Blazhko stars. Red squares represent known Blazhko 
   			stars taken from the on-line database BlaSGalF \citep{skarka2013b}. Recently discovered Blazhko stars are plotted as 
   			blue circles. Stars with periods over 300~d are not showed for a clearer arrangement.}%
   			\label{stats}
   \end{figure}

The distribution of newly discovered Blazhko variables in $P_{\mathrm{BL}}$ vs. $P_{\mathrm{puls}}$ diagram displayed in fig. \ref{stats} is random, and no dependency between the pulsation and Blazhko periods was found, as expected. 

Twelve stars in our sample turned out to be multiple-modulated and/or with changing Blazhko effect, which is 12~\% of all studied 
Blazhko stars. Until today only a few stars have been known to show such behaviour. Again, it is very likely that compound 
modulation is more common than astronomers thought and that the number of known multiple-modulated stars will increase in future analyses.


\section{Conclusions} 
 
A period study of 321 fundamental mode galactic field RR Lyraes with brightness at maximum light higher than 12.5 mag was performed based on available data from the ASAS and/or SW database. These stars were chosen from a more extended sample consisting of 557 stars after careful pre-selection based on conditions described in sec. 2. Each light curve was cleaned and analysed individually. In addition, each frequency spectrum was scanned through careful visual inspection to avoid omitting any sign of the Blahzko effect. 
	
One hundred stars were identified and confirmed to be Blazhko variables (25 previously unknown). Eight stars, marked as Blazhko candidates, showed some ambiguous indications of modulation like characteristic scattering around maximum of phased light curve. In contrast no indication of the Blazhko behaviour was found in 18 stars, which were supposed to be modulated. However, five of these eighteen stars are certain Blazhko variables, but with too small amplitude of modulation to be detectable in ASAS or SW data.
		
The incidence rate of RRab Blazhko stars from our analysis is at least 31~\%, which is a much higher percentage than given by previous studies based on data from surveys with similar data quality. This increase would be logically expected, since more time-extended datasets were examined (in many cases simultaneously in ASAS and SW surveys). However, it is still less than the incidence rates determined in the framework of precise ground-base and space surveys. With certainty there remained undetected low-amplitude Blazhko variables, as well as stars with the shortest and the longest modulation periods, simply due to nature of the data from ASAS and SW surveys.
	
A surprising result was a very high incidence of stars with some peculiarity in their Blazhko effect -- each eighth star of our Blazhko sample showed signs of the changes in Blazhko effect or compound modulation. An interesting examples are IK Hya and RZ CVn. Both stars probably undergo changes in their pulsation and modulation characteristics. IK Hya could be another star with a few-year-long cycle analogical to four-year cycle in RR~Lyrae itself. Five of ten identified multiple-modulated stars showed the ratio between their modulation periods in ratio of small integers.   
		
Our study confirmed that classification and statistics of modulated stars according to their side peaks, which are based on low-quality data, are not reliable. It was also shown that the sensitivity of automatic procedures used to identify the Blazhko effect is insufficient -- one fourth of the Blazhko variables were identified for the first time. It is likely that the incidences and characteristics of Blazhko variables identified in other surveys \citep[e.g. in LMC based on OGLE, ][]{alcock2003} could be slightly different than proposed. Our conclusions should serve as a challenge for developing better tools for automatic data analysis, because a manual examination of	many thousand stars is impossible to perform. 

\begin{acknowledgements}
Work on the paper was supported by projects MU~MUNI/A/0968/2009 and MUNI/A/0735/2012. 
The International Variable Star Index (VSX) database, operated at AAVSO, 
Cambridge, Massachusetts, USA, was used. This research also have made 
use of VizieR catalogue access tool, CDS, Strasbourg, France. We acknowledge the WASP 
consortium, which comprises of the University of Cambridge, Keele University, University of Leicester, 
The Open University, The Queen's University Belfast, St. Andrews University, and the Isaac Newton Group. 
Funding for WASP comes from the consortium universities and from the UK's Science and Technology Facilities 
Council. I would like to thank Miloslav Zejda and S.\,N.\,de\,Villiers for their suggestions, helpful comments, and language corrections. I am very grateful to my anonymous referee, who significantly helped to improve this paper.  
\end{acknowledgements}

\end{document}